\documentclass[11pt,a4paper]{article}

\usepackage[T1]{fontenc}
\usepackage[utf8]{inputenc}
\usepackage[margin=2.4cm,top=2.5cm,bottom=2.6cm]{geometry}
\usepackage{microtype}
\usepackage{amsmath,amsthm,mathtools}
\usepackage{txfonts}
\usepackage{bm}
\usepackage{graphicx}
\usepackage{booktabs,multirow,array}
\usepackage[table]{xcolor}
\usepackage{caption,subcaption}
\usepackage{algorithm,algpseudocode}
\usepackage{tikz}
\usetikzlibrary{arrows.meta,positioning,fit,backgrounds,calc,shapes.geometric}
\usepackage{quantikz}
\usepackage{enumitem}
\usepackage[numbers,sort&compress]{natbib}
\usepackage[colorlinks=true,linkcolor=blue!50!black,citecolor=blue!50!black,urlcolor=blue!50!black]{hyperref}
\usepackage[nameinlink,capitalise]{cleveref}
\makeatletter\providecommand{\theHALG@line}{}\renewcommand{\theHALG@line}{\thealgorithm.\arabic{ALG@line}}\makeatother
\usepackage{rotating}
\usepackage{placeins}

\setlist{itemsep=2pt,topsep=4pt}

\newtheorem{proposition}{Proposition}

\theoremstyle{definition}
\newtheorem{definition}{Definition}
\theoremstyle{remark}

\newcommand{\E}{\mathbb{E}}
\newcommand{\KL}{D_{\mathrm{KL}}}
\newcommand{\Dstar}{D^{\ast}}
\newcommand{\Mhon}{\mathcal{M}}
\newcommand{\tr}{^{\mathsf{T}}}
\providecommand{\ket}[1]{\lvert #1\rangle}

\providecommand{\braket}[2]{\langle #1 \vert #2\rangle}
\newcommand{\qtrace}{Q-TRACE}
\DeclareMathOperator*{\argmax}{arg\,max}

\newcommand{\h}{h_2}

\newcommand{\NSeeds}{5}
\newcommand{\SeedList}{101, 102, 103, 104, 105}
\newcommand{\NRunsTotal}{3{,}275}

\newcommand{\NWindowsTotal}{655{,}000}
\newcommand{\WallPerSeed}{13}
\newcommand{\DetKnownQT}{63.6}
\newcommand{\DetKnownBest}{51.2}
\newcommand{\DetKnownBestName}{supervised XGBoost}
\newcommand{\DetZeroDayQT}{28.0}
\newcommand{\DetZeroDayBest}{6.7}
\newcommand{\DetZeroDayBestName}{quantum-kernel one-class SVM}
\newcommand{\DetCompQT}{65.3}
\newcommand{\DetCompBest}{42.3}
\newcommand{\DetCompBestName}{supervised XGBoost}
\newcommand{\DetShiftQT}{32.0}

\newcommand{\FAHonestQT}{5.2}

\newcommand{\FAHonestQKoneclassSVM}{10.4}
\newcommand{\FALossQT}{0.0}

\newcommand{\FALossAutoencoder}{4.0}
\newcommand{\FALossIsolationforest}{12.0}

\newcommand{\FALossQKoneclassSVM}{12.8}
\newcommand{\FALongQT}{3.2}

\newcommand{\FALongQBERstatic}{45.6}

\newcommand{\DetZeroDayXGBoostsup}{0.6}

\newcommand{\DetKnownDecoyCUSUM}{2.5}

\newcommand{\DetShiftQBERstatic}{50.7}

\newcommand{\AucKnownQT}{0.859}
\newcommand{\AucZeroDayQT}{0.636}
\newcommand{\AucCompQT}{0.858}
\newcommand{\AucShiftQT}{0.699}

\newcommand{\AucShiftXGB}{0.718}
\newcommand{\AucKnownQK}{0.523}

\newcommand{\FamInterceptresendQT}{80.6}

\newcommand{\FamInterceptresendQC}{78.3}
\newcommand{\FamPNSQT}{62.9}
\newcommand{\FamPNSXGB}{46.9}
\newcommand{\FamPNSQC}{1.7}
\newcommand{\FamBlindingQT}{47.4}
\newcommand{\FamBlindingXGB}{26.3}

\newcommand{\FamTimeshiftQT}{24.0}

\newcommand{\FamTrojanhorseQT}{2.9}

\newcommand{\FamRNGbiasQT}{57.1}

\newcommand{\BoundLtOneQT}{93.7}
\newcommand{\BoundLtOneXGB}{62.0}
\newcommand{\BoundLtOneQC}{31.7}

\newcommand{\BoundOneTenQT}{25.4}
\newcommand{\BoundOneTenXGB}{6.4}
\newcommand{\BoundOneTenQC}{3.4}

\newcommand{\AttrPNSShare}{50}

\newcommand{\AttrBlindingShare}{54}

\newcommand{\AttrInterceptresendShare}{55}

\newcommand{\HwBackend}{ibm\_marrakesh}
\newcommand{\HwLinks}{33}
\newcommand{\HwShots}{5{,}900}
\newcommand{\HwPilotSec}{3}
\newcommand{\HwMainSec}{104}
\newcommand{\HwTotalSec}{107}
\newcommand{\HwRounds}{12.5}
\newcommand{\HwPreFAQT}{92}
\newcommand{\HwPreFAQP}{92}
\newcommand{\HwPreFAQB}{95}
\newcommand{\HwHonestN}{66}
\newcommand{\HwShiftZ}{0.32}

\newcommand{\HwShiftX}{1.86}
\newcommand{\HwShiftXPos}{33}
\newcommand{\HwPassDiff}{1.1\times 10^{-4}}
\newcommand{\HwQberCalMed}{1.13}
\newcommand{\HwQberCalMin}{0.27}
\newcommand{\HwQberCalMax}{11.99}
\newcommand{\HwPhi}{1.00}
\newcommand{\HwPhiQ}{1.014}
\newcommand{\HwThetaAQT}{14}
\newcommand{\HwThetaAQB}{13}
\newcommand{\HwThetaAQP}{9}
\newcommand{\HwThetaAN}{14}

\newcommand{\HwThetaAIdeal}{0.28}

\newcommand{\HwThetaDMeas}{4.34}
\newcommand{\HwThetaDIdeal}{4.37}

\newcommand{\HwThetaEMeas}{9.05}
\newcommand{\HwThetaEIdeal}{9.46}

\newcommand{\HwShotsWeakQT}{511}
\newcommand{\HwLeakWeakQT}{14.9}
\newcommand{\HwShotsWeakQB}{432}
\newcommand{\HwLeakWeakQB}{12.6}
\newcommand{\HwShotsWeakQP}{2851}
\newcommand{\HwLeakWeakQP}{83.0}
\newcommand{\HwShotsWeakOM}{645}

\newcommand{\HwShotsRatio}{5.6}
\newcommand{\HwIRInfo}{0.699}
\newcommand{\HwIRGuess}{0.053}
\newcommand{\HwInfoFourMeas}{0.090}
\newcommand{\HwInfoFourIdeal}{0.112}

\title{Q-TRACE: Physics-based zero-day attack detection for decoy-state quantum key distribution}
\author{Sumit Tapas Chongder\\[4pt]
\small Department of Computer Science \& Engineering and Department of Physics,\\
\small Indian Institute of Technology Jodhpur, Jodhpur, Rajasthan, India}
\date{}

\begin{document}
\maketitle

\begin{abstract}
\noindent
Practical quantum key distribution (QKD) is secure only as long as its devices behave as the security proof assumes, and every published implementation attack works by violating one of those assumptions while keeping the quantum bit error rate (QBER) unremarkable. Telemetry monitors are a natural complement to device countermeasures, but the data-driven monitors proposed so far are trained on simulated attacks, offer no false-alarm guarantee, and cannot say which attacks they are able to see. We formulate attack detection in decoy-state BB84 as a sequential test against the \emph{honest-device manifold}: the set of receiver outcome statistics that an attack-free link can produce under arbitrary channel loss. The resulting monitor, \qtrace{} (Quantum Telemetry Residuals Against Calibrated Expectations), calibrates a physical model of each link on commissioning data, treats transmittance as a nuisance parameter, and combines an omnibus residual test, a physics score test along the directions in which device parameters can move, and a setting-allocation test, each with a drift-robust moving reference. No attack data enter its construction. The same formulation yields a detector-independent detectability exponent $\Dstar=\min_\eta \KL(Q\Vert P_\eta)$ which, through Lorden's theorem, lower-bounds the delay of any telemetry monitor, and which vanishes identically for a Trojan-horse probe with filtered back-reflection. On a count-level twin of a decoy-state link that contains detector, source and drift effects absent from the monitor's model, and over \NSeeds{} held-out seeds, \qtrace{} detects \DetKnownQT\,\% of links under known attacks, \DetZeroDayQT\,\% under attacks of families never used in development, and \DetCompQT\,\% under composite attacks at a per-window false-alarm rate of $10^{-3}$; the strongest comparison methods reach \DetKnownBest\,\%, \DetZeroDayBest\,\% and \DetCompBest\,\%. Its median detection delay attains the information-theoretic bound for strongly detectable attacks and stays within a small factor of it up to bounds of about ten windows, whereas QBER and learned monitors need one to two orders of magnitude more data for the same attacks. A hardware study on \HwLinks{} parallel links of the IBM \HwBackend{} processor, with eavesdroppers executed as tunable entangling and amplitude-damping probes, reproduces the ideal information--disturbance relations, shows that a commissioning reference acquired under different operating conditions triggers alarms in every monitor tested, and, with a condition-matched reference, detects a probe with an ideal QBER increase of $0.3$\,\% on all \HwThetaAN{} attacked link-epochs with no false alarm on \HwHonestN{} honest ones.
\end{abstract}


\section{Introduction}\label{sec:intro}

Quantum key distribution allows two parties to establish a shared secret whose security rests on quantum mechanics rather than on computational assumptions \cite{BB84,Ekert1991,ShorPreskill2000}. Two decades of work have turned the protocols into deployed systems, including metropolitan networks and satellite links \cite{Liao2017,Chen2021}, and the security of realistic implementations with attenuated laser sources is now well understood \cite{GLLP2004,Hwang2003,LoMaChen2005,Wang2005,Ma2005,Scarani2009,Xu2020,Pirandola2020}. Decoy-state BB84 in particular has become the workhorse of practical QKD.

The security of a real system, however, is only as good as the match between its devices and the model used in the proof. Every implementation attack in the literature exploits a gap of this kind: photon-number splitting exploits multi-photon pulses \cite{Brassard2000,Lutkenhaus2000}; detector blinding turns single-photon avalanche diodes into classical threshold devices that the eavesdropper controls \cite{Lydersen2010,Gerhardt2011,Wiechers2011,Lydersen2011superlinear}; time-shift and efficiency-mismatch attacks exploit detectors whose efficiencies differ in time \cite{Makarov2006,Qi2007,Zhao2008}; Trojan-horse attacks read modulator settings through back-reflected light \cite{Gisin2006THA,Jain2014THA,Lucamarini2015}; and calibration, phase-remapping, laser-damage and laser-seeding attacks manipulate the devices themselves \cite{Jain2011calibration,XuQiLo2010,Bugge2014,Makarov2016,Huang2019seeding}. Reviews of this landscape stress that the QBER, the quantity a system monitors by default, is precisely what a competent eavesdropper keeps low \cite{Jain2016review,SunHuang2022,Diamanti2016}.

Three lines of defence exist. Device-independent and measurement-device-independent protocols remove whole classes of assumptions \cite{Acin2007,LoCurtyQi2012} at a cost in rate and complexity. Hardware countermeasures close individual side channels \cite{Lucamarini2015,Ponosova2022}, but each closes only the channel it was designed for and may open others. Security proofs can be extended to cover characterised imperfections \cite{Tamaki2016,Pereira2020,Zapatero2021}, which presupposes that the imperfection is known. A fourth, complementary line is \emph{monitoring}: the legitimate parties already share rich statistics after sifting and error correction, and an attack that changes those statistics can, in principle, be noticed.

Monitoring has recently attracted machine-learning approaches. Supervised and semi-supervised neural networks have been trained to recognise attacks on continuous-variable systems \cite{Mao2020,Luo2022} and on discrete-variable systems \cite{XuJ2024,AlMohammed2021,Ren2021}; one-class anomaly detectors have been applied to receiver attacks on an experimental system \cite{Liu2025AD}; hybrid quantum--classical recurrent networks have been proposed for decoy-state BB84 telemetry \cite{AlKuwari2025}; and adversarial training against adaptive attackers has been explored \cite{MohamedAlKuwari2026}. These studies show that attacks leave learnable traces. They also share three limitations that matter for deployment. First, supervised detectors are trained on the attacks they are tested on, which is not the situation a system faces against an unknown attack. Second, the false-alarm rate is an empirical property of a particular test set rather than a controlled design parameter, although a false alarm discards key and a missed alarm compromises it. Third, none of these works asks which attacks are detectable \emph{at all} from the available telemetry, so a poor result cannot be distinguished from a fundamentally invisible attack.

This paper addresses the three limitations with a single idea. Instead of learning what attacks look like, we characterise what an honest link can look like. Given the receiver statistics of a decoy-state link, the set of distributions consistent with \emph{some} attack-free channel, with transmittance left free because loss is accounted for by the decoy-state bound, forms a low-dimensional manifold. Detection becomes a composite hypothesis test against that manifold, and the information divergence from an attacked distribution to the manifold measures how detectable the attack is for any monitor whatsoever.

\paragraph{Contributions.} The work makes four contributions.
\begin{enumerate}[label=(\roman*)]
\item \emph{An honest-manifold monitor.} We derive a count-level model of a two-detector decoy-state receiver that is exact for threshold detectors and photon-number-dependent channels (\cref{sec:model}), and build on it a sequential monitor, which we call \qtrace{} (Quantum Telemetry Residuals Against Calibrated Expectations), that calibrates each link on commissioning data and combines an omnibus residual test, a physics score test with an efficient treatment of the transmittance nuisance, and a setting-allocation test, each with a drift-robust moving reference (\cref{sec:detector}). The monitor is never shown an attack.
\item \emph{Detectability limits.} We define the detectability exponent $\Dstar$, show that it lower-bounds the delay of every telemetry monitor at a given false-alarm rate, give its local form in terms of the efficient Fisher information, and prove that a Trojan-horse probe with filtered back-reflection is invisible to all telemetry monitors (\cref{sec:theory}).
\item \emph{A benchmark with model mismatch.} We construct six physically derived attacks and three compositions on a count-level twin that contains afterpulsing, intensity jitter, polarisation and temperature drift, none of which the monitor models, and compare \qtrace{} with industry monitors, four anomaly detectors including a quantum-kernel method, and a supervised classifier under matched false-alarm rates on held-out seeds (\cref{sec:benchmark,sec:results}).
\item \emph{A hardware study.} We run \HwLinks{} BB84 links in parallel on an IBM Heron processor with eavesdroppers implemented as tunable two-qubit probes, measure the eavesdropper's information on her ancilla, and report both a pre-registered analysis, in which every monitor raised alarms on honest links because the commissioning reference had been acquired under different operating conditions, and a condition-matched analysis (\cref{sec:hardware}).
\end{enumerate}
The code, the raw hardware data and every number in this paper are available in the accompanying repository; each quantitative statement is generated from the result files by a script.

\paragraph{What this work does not claim.} The monitor complements security proofs and device countermeasures; it does not replace them. Attacks with $\Dstar=0$ cannot be detected from telemetry by any method, and we identify one. The simulation results depend on the stated parameter ranges of the twin, and the hardware study emulates the quantum channel with superconducting qubits rather than optical fibre.

\paragraph{Outline.} \Cref{sec:background} reviews decoy-state BB84, implementation attacks and change detection. \Cref{sec:model} derives the receiver model and \cref{sec:attacks} the attack models. \Cref{sec:detector} constructs the monitor and \cref{sec:theory} develops the detectability theory. \Cref{sec:benchmark,sec:results} describe the benchmark and its results, \cref{sec:hardware} the hardware study, and \cref{sec:discussion} discusses implications and limitations. Derivations, parameter tables and supplementary results are collected in the appendices.

\section{Background}\label{sec:background}

\subsection{Decoy-state BB84}\label{sec:bg-decoy}

In BB84 \cite{BB84} Alice encodes a random bit $a$ in one of two conjugate bases, $\alpha\in\{Z,X\}$, and Bob measures in a basis $\beta$ chosen independently; rounds with $\alpha=\beta$ form the sifted key. With a phase-randomised weak coherent source the number of photons in a pulse of mean $\mu$ is Poisson distributed, so a fraction $1-e^{-\mu}(1+\mu)$ of pulses carries two or more photons, and an eavesdropper able to split those photons learns their bits without disturbance \cite{Brassard2000,Lutkenhaus2000}. The GLLP analysis \cite{GLLP2004} shows that secure key can still be extracted from the single-photon component; the decoy-state method \cite{Hwang2003,LoMaChen2005,Wang2005} estimates the single-photon yield $Y_1$ and error rate $e_1$ by sending pulses of several intensities that the eavesdropper cannot distinguish. With a signal intensity $\mu$, a weak decoy $\nu<\mu$ and a vacuum decoy, the bounds of Ma \emph{et al.} \cite{Ma2005} read
\begin{align}
Y_1 &\ge Y_1^{L} = \frac{\mu}{\mu\nu-\nu^2}\left(Q_\nu e^{\nu} - Q_\mu e^{\mu}\frac{\nu^2}{\mu^2} - \frac{\mu^2-\nu^2}{\mu^2}Y_0\right), \label{eq:y1}\\
e_1 &\le e_1^{U} = \frac{E_\nu Q_\nu e^{\nu} - e_0 Y_0}{Y_1^{L}\,\nu}, \label{eq:e1}
\end{align}
where $Q_k$ and $E_k$ are the gain and QBER of intensity $k$, $Y_0$ is the vacuum yield and $e_0=1/2$. The asymptotic key rate per signal pulse is
\begin{equation}
R = Q_1\bigl[1-\h(e_1^{U})\bigr] - f\,Q_\mu \h(E_\mu),\qquad Q_1 = Y_1^{L}\mu e^{-\mu},\label{eq:rate}
\end{equation}
with $\h$ the binary entropy and $f\ge 1$ the error-correction inefficiency. We use the efficient variant in which key is distilled from $Z$-basis rounds and the phase error is bounded with $X$-basis decoy statistics \cite{LoChauArdehali2005}. Finite-size corrections \cite{Tomamichel2012,Lim2014} are not needed for the monitoring problem and are not used here.

The structure of \cref{eq:y1,eq:e1,eq:rate} is the key physical fact behind this paper: the decoy-state bound is valid for \emph{any} channel transmittance, so a change of loss alone is not a security-relevant event. What an attack cannot avoid without restriction is a change in the relations between intensities, bases, bit values and detectors that an honest lossy channel would impose.

\subsection{Implementation attacks}\label{sec:bg-attacks}

\Cref{tab:attacks} summarises the attacks considered in this work and the telemetry signatures they produce. Intercept--resend raises the QBER by a quarter of the intercepted fraction. Photon-number splitting with a lossless channel \cite{Brassard2000} preserves the signal gain but breaks the decoy relation. Faked-state attacks on blinded detectors \cite{Lydersen2010,Gerhardt2011} can reproduce the gain and even the QBER, but blinded detectors produce neither dark counts nor double clicks. Time-shift attacks \cite{Qi2007,Zhao2008} exploit efficiency mismatch \cite{Makarov2006} and change the error structure in a bit-dependent way. Trojan-horse attacks \cite{Gisin2006THA,Lucamarini2015} act on Alice and need not affect Bob at all. A partially predictable random-number generator changes only the setting statistics. These six families span passive and active attacks, transmitter and receiver, and photon-number, timing, detector and randomness side channels.

\begin{table}[!tbp]
\centering
\caption{Attack families used in this work, their physical mechanism and their signature in receiver telemetry. ``Known'' families are available to the supervised baseline during training; ``zero-day'' families are used only for testing and were never used in the design of \qtrace{}.}
\label{tab:attacks}
\small
\begin{tabular}{@{}>{\raggedright\arraybackslash}p{2.6cm}>{\raggedright\arraybackslash}p{1.35cm}>{\raggedright\arraybackslash}p{5.6cm}>{\raggedright\arraybackslash}p{5.2cm}@{}}
\toprule
Family & Role & Mechanism & Telemetry signature \\
\midrule
Intercept--resend & known & Measure in random basis, resend found state & Sifted error $e\to e/2+1/4$ on attacked pulses \\
Photon-number splitting & known & Keep one photon of multi-photon pulses, lossless forwarding, suppress single photons to match gain \cite{Brassard2000} & Decoy/signal gain relation, double-click rate \\
Detector blinding & known & Faked states on blinded detectors, gain and QBER matched \cite{Lydersen2010} & No dark counts, no double clicks, unsifted balance $\to 1/2$ \\
Time-shift & zero-day & Exploit efficiency mismatch; loss compensated \cite{Qi2007,Zhao2008} & Bit-dependent errors, double clicks \\
Trojan horse & zero-day & Probe Alice's modulators \cite{Gisin2006THA}; optional back-scatter to Bob & Background counts only; none if filtered \\
RNG bias & zero-day & Bit generator modulated as $\tfrac12+\delta\sin(2\pi t/T)$ & Setting allocation only \\
\bottomrule
\end{tabular}
\end{table}

\subsection{Sequential change detection}\label{sec:bg-cd}

A monitor observes a sequence of windows and must raise an alarm soon after a change of distribution while rarely alarming without one. For i.i.d.\ observations with known pre- and post-change laws $P$ and $Q$, Page's CUSUM \cite{Page1954} is optimal in Lorden's minimax sense \cite{Lorden1971,Moustakides1986}, and every procedure whose average run length to false alarm is at least $\gamma$ has a worst-case mean delay of at least $(1+o(1))\log\gamma/\KL(Q\Vert P)$ as $\gamma\to\infty$ \cite{Lorden1971,Lai1998}. When the pre-change law contains unknown parameters and the post-change law is unknown, generalised likelihood ratio (GLR) and score procedures are used \cite{Basseville1993,Tartakovsky2014}. The score (Rao) test \cite{Rao1948} and Neyman's $C(\alpha)$ construction \cite{Neyman1959} are locally most powerful against alternatives along chosen directions and handle nuisance parameters through the efficient score. These tools, standard in industrial process control, have not, to our knowledge, been used to build attack monitors for QKD from a physical model of the receiver.

\section{A count-level model of the decoy-state receiver}\label{sec:model}

\subsection{Settings, outcomes and telemetry}\label{sec:settings}

Each transmitted pulse carries a \emph{setting} $s=(k,a,\alpha,\beta)$: the intensity class $k\in\{\mathrm{s},\mathrm{d},\mathrm{v}\}$ (signal, decoy, vacuum) with means $\mu_\mathrm{s}>\mu_\mathrm{d}>\mu_\mathrm{v}=0$, Alice's bit $a\in\{0,1\}$, Alice's basis $\alpha$ and Bob's basis $\beta$. Settings are drawn independently with probabilities
\begin{equation}
\pi_s = p_k\,p_a\,p_\alpha\,p_\beta, \label{eq:settingprob}
\end{equation}
so there are $3\times2\times2\times2=24$ settings. Bob's passive receiver has two threshold detectors, $D_0$ and $D_1$, encoding bit values; a pulse produces one of four \emph{outcomes} $o\in\{\varnothing,D_0,D_1,D_{01}\}$ (no click, a click on one detector, or a double click). After sifting and error correction Alice and Bob know, for each telemetry window of $N$ pulses, the full table of counts $n_{s,o}$, 96 numbers in total. The error count of the sifted key is revealed exactly by error correction, and the outcomes of decoy, vacuum and basis-mismatched rounds are discarded from the key anyway, so publishing them costs no key. Double clicks are assigned a uniformly random bit (squashing), which is the standard convention in security proofs.

Conditioned on the allocation $N_s=\sum_o n_{s,o}$, the outcome counts of setting $s$ are multinomial,
\begin{equation}
(n_{s,o})_o \sim \mathrm{Mult}\bigl(N_s;\,(p_{o\mid s})_o\bigr), \qquad (N_s)_s \sim \mathrm{Mult}\bigl(N;\,(\pi_s)_s\bigr). \label{eq:mult}
\end{equation}
The log-likelihood of a window therefore separates into an outcome part and an allocation part,
\begin{equation}
\ell = \sum_{s,o} n_{s,o}\log p_{o\mid s} + \sum_s N_s \log \pi_s + \text{const}, \label{eq:loglik}
\end{equation}
and the monitor of \cref{sec:detector} uses both.

\subsection{Threshold detectors and photon-number-dependent channels}\label{sec:threshold}

Let $S_j$ denote the probability that detector $j$ stays silent and $S_{01}$ the probability that both stay silent. Inclusion--exclusion gives the four outcome probabilities
\begin{equation}
p_{\varnothing}=S_{01},\quad p_{D_0}=S_1-S_{01},\quad p_{D_1}=S_0-S_{01},\quad p_{D_{01}}=1-S_0-S_1+S_{01}. \label{eq:outcomes}
\end{equation}
To compute the silence probabilities, suppose that $n$ photons leave Alice and $n'$ of them reach Bob. Each arriving photon is routed to detector $j$ and fires it with probability $\varpi_j$, independently of the others, where
\begin{equation}
\varpi_j(a,\alpha,\beta) = \eta_B\, w_j(a,\alpha,\beta)\, d_j,\qquad
w_j = \begin{cases} 1-e_\beta & j=a,\ \alpha=\beta,\\ e_\beta & j\neq a,\ \alpha=\beta,\\ 1/2 & \alpha\neq\beta,\end{cases} \label{eq:routing}
\end{equation}
with $\eta_B$ the receiver efficiency, $e_\beta$ the optical misalignment of basis $\beta$, $d_0=1$ and $d_1=\kappa$ the efficiency of $D_1$ relative to $D_0$. Firing events of different photons at the same detector are not mutually exclusive, but routing of a single photon is; hence, given $n'$ arriving photons,
\begin{equation}
\Pr(D_j\ \text{silent}\mid n') = (1-\varpi_j)^{n'},\qquad \Pr(\text{both silent}\mid n') = (1-\varpi_0-\varpi_1)^{n'}.
\end{equation}
Background clicks (dark counts) occur independently with probability $p_{dj}$ per gate. Averaging over the channel with the \emph{arrival generating function} $G_n(z)=\E[z^{n'}\mid n]$ and over the photon-number distribution $P(n)$ of the source yields the exact expressions
\begin{equation}
S_j = (1-p_{dj})\sum_{n\ge0}P(n)\,G_n(1-\varpi_j),\qquad
S_{01} = (1-p_{d0})(1-p_{d1})\sum_{n\ge0}P(n)\,G_n(1-\varpi_0-\varpi_1). \label{eq:silence}
\end{equation}
\Cref{eq:outcomes,eq:routing,eq:silence} hold for any channel that acts on photon numbers, including the photon-number-splitting channel of \cref{sec:attacks}.

\begin{proposition}[Poissonian source, lossy channel]\label{prop:poisson}
For a phase-randomised coherent source of mean $\mu$, $P(n)=e^{-\mu}\mu^n/n!$, and a memoryless channel of transmittance $t$, $G_n(z)=(1-t+tz)^n$, and
\begin{equation}
S_j=(1-p_{dj})\exp\!\bigl(-\mu\eta\, w_j d_j\bigr),\qquad S_{01}=S_0\,S_1, \label{eq:poisson}
\end{equation}
with $\eta=t\eta_B$; in particular the two detectors fire independently.
\end{proposition}
\begin{proof}
$\sum_n e^{-\mu}\mu^n(1-t+tz)^n/n! = \exp[-\mu t(1-z)]$. Substituting $z=1-\varpi_j$ gives $\exp(-\mu t\varpi_j)=\exp(-\mu\eta w_jd_j)$, and $z=1-\varpi_0-\varpi_1$ gives the product of the two single-detector factors. This is the binomial thinning property of the Poisson distribution.
\end{proof}

The honest model is completed by detector afterpulsing, which adds to each detector a background proportional to its mean click probability. With $\bar r_j=\sum_s\pi_s\Pr(D_j\ \text{fires}\mid s)$ and afterpulse probability $p_{\rm ap}$, the extra click probability is $q_j=p_{\rm ap}\bar r_j$, and the silence probabilities become $S_j\to S_j(1-q_j)$ and $S_{01}\to S_{01}(1-q_0)(1-q_1)$. We verified numerically that the generating-function route of \cref{eq:silence} and the closed form of \cref{eq:poisson} agree to $2\times10^{-16}$.

\subsection{The honest-device family}\label{sec:family}

For an attack-free link, \cref{eq:outcomes,eq:routing,eq:poisson} define the outcome probabilities as a function of
\begin{equation}
\psi=(e_Z,\,e_X,\,p_{d0},\,p_{d1},\,\kappa,\,\rho)\quad\text{and}\quad \eta, \label{eq:params}
\end{equation}
where $\rho=\mu_\mathrm{d}/\mu_\mathrm{s}$ is the decoy-to-signal intensity ratio. The signal intensity $\mu_\mathrm{s}$ is read from the transmitter's power monitor and is in any case degenerate with $\eta$; the afterpulse probability is taken from the detector specification. We write $P_{\psi,\eta}$ for the resulting distribution over settings and outcomes. For a link whose device parameters $\psi$ have been calibrated, the \emph{honest-device manifold} is the one-parameter family
\begin{equation}
\Mhon_\psi = \{P_{\psi,\eta}:\ \eta\in(0,1]\}. \label{eq:manifold}
\end{equation}
Treating $\eta$ as free encodes the physics of \cref{sec:bg-decoy}: changes of loss are absorbed by the decoy-state bound and should not be reported as attacks. Everything that no transmittance can explain, a broken decoy relation, a changed error structure, an anomaly of detector balance or double clicks, a biased setting allocation, is a candidate attack signature.

\section{Physical models of the attacks}\label{sec:attacks}

Each attack acts on a fraction $f$ of the pulses from its onset window onwards, so that the attacked distribution is $Q=(1-f)P_{\psi,\eta}+fP^{\rm atk}$ for the outcome part and analogously for the allocation part. We derive $P^{\rm atk}$ for each family from its physical mechanism; the derivations are written at the level of the per-setting outcome probabilities so that they enter the same likelihood as the honest model.

\paragraph{Intercept--resend.} Eve measures in a basis chosen uniformly at random and resends the state she found, preserving the photon statistics. For matching bases Bob's error is $e_\beta$ when Eve chose Alice's basis and $1/2$ otherwise, so
\begin{equation}
e_\beta \;\to\; e'_\beta = \tfrac12 e_\beta + \tfrac14. \label{eq:ir}
\end{equation}
The sifted QBER of attacked pulses rises by $\tfrac14(1-2e_\beta)$, which is the textbook signature.

\paragraph{Photon-number splitting.} Eve replaces the channel by a lossless one, stores one photon of every multi-photon pulse and suppresses single photons so that Bob's signal gain is unchanged \cite{Brassard2000}. In the generating-function language,
\begin{equation}
G_0(z)=1,\qquad G_1(z)=1-t_1+t_1z,\qquad G_{n\ge2}(z)=(1-b)\,z^{\,n-1}+b, \label{eq:pns}
\end{equation}
where $t_1$ is the transmitted fraction of single-photon pulses and $b$ the blocked fraction of multi-photon pulses. Eve chooses $(t_1,b)$ so that the signal click probability equals its honest value $G_\mathrm{s}$: if forwarding all multi-photon pulses already exceeds $G_\mathrm{s}$ she sets $t_1=0$ and solves for $b$, otherwise $b=0$ and she solves for $t_1$. Both conditions are monotone and have a unique solution, found by bisection. Because the decoy intensity has a smaller multi-photon fraction, the decoy gain no longer obeys the honest relation; this is the effect the decoy method was designed to expose \cite{Hwang2003,LoMaChen2005}.

\paragraph{Faked-state attack on blinded detectors.} With Bob's detectors blinded \cite{Lydersen2010}, Eve detects Alice's pulse with efficiency $\eta_E$, chooses a basis $\gamma$ uniformly, and resends a bright trigger encoding her result. A blinded detector clicks deterministically when Bob's basis equals $\gamma$ and not at all otherwise; it produces neither dark counts nor double clicks. With $P_E(k)=1-e^{-\mu_k\eta_E}$ and Eve's own error rate $e_E$, the attacked outcome probabilities are
\begin{equation}
p^{\rm atk}_{D_a\mid s}=\tfrac12P_E(k)(1-e_E),\quad p^{\rm atk}_{D_{1-a}\mid s}=\tfrac12P_E(k)e_E\quad(\alpha=\beta),\qquad p^{\rm atk}_{D_0\mid s}=p^{\rm atk}_{D_1\mid s}=\tfrac14P_E(k)\quad(\alpha\ne\beta), \label{eq:blind}
\end{equation}
with $p^{\rm atk}_{D_{01}\mid s}=0$ and $p^{\rm atk}_{\varnothing\mid s}=1-\tfrac12P_E(k)$. A stealthy Eve matches the signal gain, $\eta_E=-\ln(1-2G_\mathrm{s})/\mu_\mathrm{s}$, and the honest QBER, $e_E=E_\mathrm{s}$. For small $\mu\eta_E$ the gains of all intensities scale linearly, so the decoy relation is nearly preserved; the remaining signatures are a vanishing vacuum yield, the absence of double clicks and an unsifted detector balance of exactly $1/2$ irrespective of $\kappa$.

\paragraph{Time-shift attack.} Each attacked pulse is shifted to a time at which one detector's efficiency is reduced by a factor $r<1$ \cite{Qi2007,Zhao2008}; with equal probability the favoured detector is $D_0$ or $D_1$. Eve compensates the loss of gain by lowering the channel loss by a factor $g=\min\{2/(1+r),\,1/t\}$. The attacked distribution is the even mixture of two honest-form distributions with transmittance $g\eta$ and relative efficiencies $(1,\kappa r)$ and $(r,\kappa/r)$ respectively, and the QBER of bit $a$ becomes asymmetric because errors are routed to the disfavoured detector half the time.

\paragraph{Trojan horse.} The probe reads Alice's modulators \cite{Gisin2006THA,Lucamarini2015} and is invisible to Bob except through back-scattered light reaching his detectors, modelled as a background $q_{\rm bs}$ added to both detectors. Information leaked to Eve is priced into the key rate by the bounds of Lucamarini \emph{et al.} \cite{Lucamarini2015}; for monitoring, the relevant fact is that with $q_{\rm bs}=0$ the attacked receiver statistics are \emph{identical} to the honest ones (\cref{prop:blind}).

\paragraph{Random-number-generator bias.} A partially predictable bit generator at Alice is modelled as $p_a(1)=\tfrac12+\delta\sin(2\pi t/T+\phi)$, which changes the allocation part of \cref{eq:loglik} and nothing else. The sinusoidal form, whose time average is exactly balanced, deliberately defeats monitors that average the bit balance over long periods.

\paragraph{Compositions.} Composite attacks superpose the first-order perturbations of their components, $Q=P+\sum_i f_i(P^{\rm atk}_i-P)$, which is exact when the components act on disjoint subsets of pulses.

\section{The \qtrace{} monitor}\label{sec:detector}

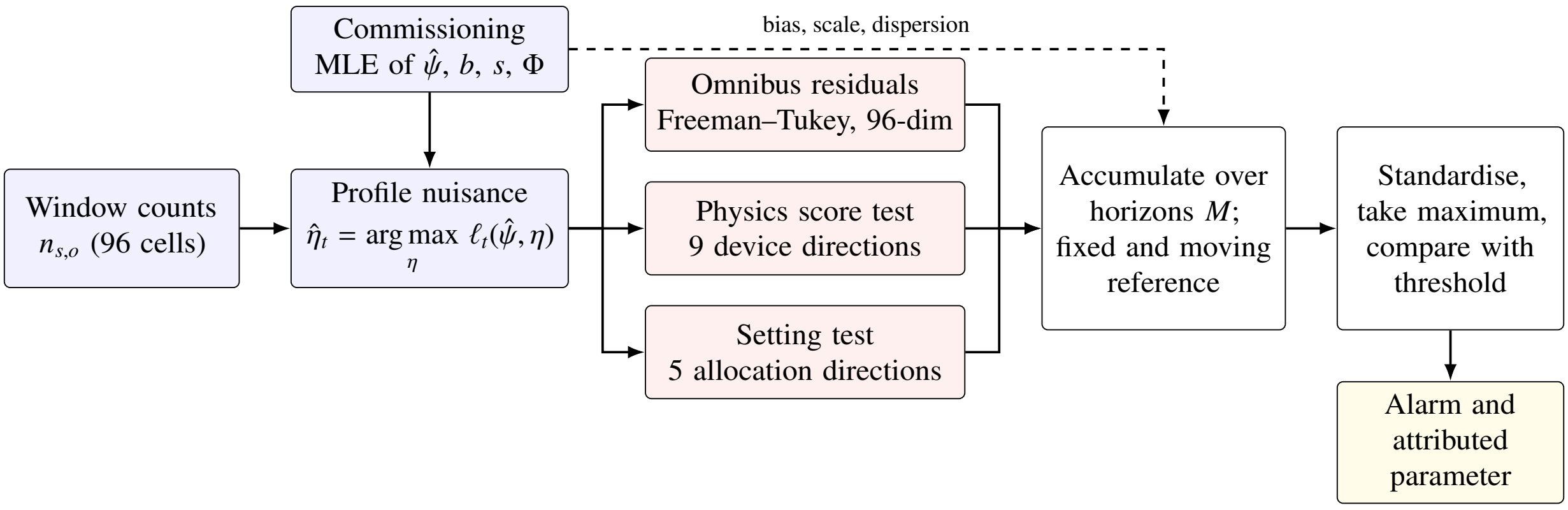
\begin{figure}[!tbp]
\centering
\resizebox{\textwidth}{!}{%
\begin{tikzpicture}[
  font=\small,
  box/.style={draw, rounded corners=2pt, align=center, inner sep=4pt, fill=white},
  hb/.style={box, fill=blue!6},
  cb/.style={box, fill=red!6, text width=3.5cm, minimum height=11mm},
  arr/.style={-{Latex[length=2mm]}, thick},
  lab/.style={font=\scriptsize, align=center}
]
\node[hb, text width=2.5cm, minimum height=14mm] (tab) {Window counts\\ $n_{s,o}$ (96 cells)};
\node[hb, text width=3.0cm, minimum height=14mm, right=0.6cm of tab] (prof) {Profile nuisance\\[2pt] $\hat\eta_t=\argmax\limits_{\eta}\,\ell_t(\hat\psi,\eta)$};
\node[hb, text width=3.0cm, above=0.9cm of prof] (cal) {Commissioning\\ MLE of $\hat\psi$, $b$, $s$, $\Phi$};
\node[cb, right=0.9cm of prof] (score) {Physics score test\\ 9 device directions};
\node[cb, above=0.35cm of score] (omni) {Omnibus residuals\\ Freeman--Tukey, 96-dim};
\node[cb, below=0.35cm of score] (set) {Setting test\\ 5 allocation directions};
\node[box, text width=2.6cm, minimum height=24mm, right=0.9cm of score] (acc) {Accumulate over\\ horizons $M$;\\ fixed and moving\\ reference};
\node[box, text width=2.4cm, minimum height=24mm, right=0.6cm of acc] (max) {Standardise,\\ take maximum,\\ compare with\\ threshold};
\node[box, text width=2.4cm, below=0.6cm of max, fill=yellow!10] (attr) {Alarm and\\ attributed\\ parameter};
\draw[arr] (tab) -- (prof);
\draw[arr] (cal) -- (prof);
\draw[arr] (prof.east) -- ++(0.4,0) |- (omni.west);
\draw[arr] (prof) -- (score);
\draw[arr] (prof.east) -- ++(0.4,0) |- (set.west);
\draw[arr] (omni.east) -- ++(0.4,0) |- (acc.west);
\draw[arr] (score) -- (acc);
\draw[arr] (set.east) -- ++(0.4,0) |- (acc.west);
\draw[arr] (acc) -- (max);
\draw[arr] (max) -- (attr);
\draw[arr, dashed] (cal.east) -| node[lab, above, pos=0.25] {bias, scale, dispersion} (acc.north);
\end{tikzpicture}}
\caption{Structure of the \qtrace{} monitor. Commissioning windows (top) fix the device parameters $\hat\psi$ and the static corrections; in every monitoring window the transmittance is profiled out and three families of statistics measure what no honest transmittance can explain. No attack data are used in any block.}
\label{fig:pipeline}
\end{figure}

\Cref{fig:pipeline} summarises the construction. The monitor is defined entirely by the honest model of \cref{sec:model} and by the link's own commissioning windows; thresholds are set on attack-free validation links. We describe its components in turn.

\subsection{Calibration and profiling}\label{sec:calibration}

Let $\ell_t(\psi,\eta)$ be the log-likelihood \eqref{eq:loglik} of window $t$. On $W$ commissioning windows the device parameters are estimated by maximising the profile likelihood
\begin{equation}
\hat\psi = \argmax_\psi \sum_{t=1}^{W} \max_{\eta_t}\ \ell_t(\psi,\eta_t). \label{eq:calib}
\end{equation}
We parametrise $\psi$ on unconstrained scales (logit for $e_Z,e_X$, logarithm for the others), start from closed-form moment estimates (vacuum click rates for $p_{dj}$, sifted signal QBER for $e_\beta$, unsifted detector ratio for $\kappa$, ratio of log-gains for $\rho$) and use a quasi-Newton optimiser. The inner maximisation over $x=\log\eta_t$ is one-dimensional and smooth and is solved for all windows simultaneously by Newton iterations,
\begin{equation}
x^{(i+1)} = x^{(i)} - \frac{\partial_x\ell_t}{\partial_x^2\ell_t}\bigg|_{x^{(i)}}, \label{eq:newton}
\end{equation}
with steps clipped to $\pm0.5$ and finite-difference derivatives. On a single CPU core a calibration takes 0.3 to 0.5\,s per link. In monitoring windows the same profiling gives $\hat\eta_t$, the closest honest explanation of window $t$ in transmittance. \Cref{alg:calib} lists the procedure.

\begin{algorithm}[t]
\caption{Commissioning of one link}\label{alg:calib}
\begin{algorithmic}[1]
\Require counts $n^{(t)}_{s,o}$ of $W$ attack-free windows, $\mu_\mathrm{s}$, datasheet $p_{\rm ap}$, setting probabilities $\pi$
\State $\psi^{(0)}\gets$ moment estimates from vacuum, sifted and unsifted counts
\State $\hat\psi\gets\argmax_\psi\sum_t\max_{\eta_t}\ell_t(\psi,\eta_t)$ \Comment{quasi-Newton outside, \cref{eq:newton} inside}
\For{$t=1,\dots,W$}
  \State $\hat\eta_t\gets\argmax_\eta \ell_t(\hat\psi,\eta)$;\ \ $r_t\gets$ Freeman--Tukey residuals \eqref{eq:ft}
  \State $(u_t,I_t)\gets$ efficient scores and informations \eqref{eq:effscore}
\EndFor
\State $b\gets\bar r$;\ \ $s^2\gets (W\hat\sigma^2_r+\lambda)/(W+\lambda)$, floored at $s_{\min}^2$ \Comment{residual bias and spread}
\State $b_u\gets\bar u$;\ \ $\Phi\gets$ shrunk covariance of $I_t^{-1/2}(u_t-b_u)$, eigenvalues floored at 1 \Comment{\cref{eq:disp}}
\State \Return $\hat\psi,b,s,b_u,\Phi$ and the commissioning statistics (for the moving reference)
\end{algorithmic}
\end{algorithm}

\subsection{Omnibus residual channel}\label{sec:omnibus}

For each of the 72 click cells (24 settings times three click outcomes) and each of the 24 allocation cells, with expected count $E$ under the profiled honest model, we form the Freeman--Tukey residual \cite{FreemanTukey1950}
\begin{equation}
r = \sqrt{n}+\sqrt{n+1}-\sqrt{4E+1}. \label{eq:ft}
\end{equation}
For Poisson or binomial counts $r$ has mean close to zero and variance close to one even when $E$ is of order unity (\cref{app:ft}), which matters for vacuum double clicks with $E\ll1$ and for signal no-click cells with $E\sim10^6$ alike. Static mismatch between the twin and the honest model (for instance a datasheet afterpulse probability in error by ten per cent) shifts the residuals by a constant; we therefore whiten them with the commissioning statistics, $z_t=(r_t-b)/s$, where $b$ and $s$ are the per-component commissioning mean and shrunk standard deviation.

Two statistics accumulate the whitened residuals over a trailing horizon of $M$ windows. The \emph{persistent-shift} statistic
\begin{equation}
S^{\rm sh}_M(t)=\frac{1}{M(1+M/W)}\Bigl\Vert\sum_{\tau=t-M+1}^{t}z_\tau\Bigr\Vert^2 \label{eq:shift}
\end{equation}
detects deviations of fixed sign; the factor $1+M/W$ accounts for the variance added by subtracting the estimated bias $b$ (\cref{app:inflation}). The \emph{energy} statistic $S^{\rm en}_M(t)=\sum_\tau\Vert z_\tau\Vert^2$ detects deviations whose sign changes in time, such as the modulated RNG attack. Both are centred and scaled by their chi-square references, $(S-MD)/\sqrt{2MD}$ with $D=96$ for the energy statistic and $(S-D)/\sqrt{2D}$ for the shift statistic.

\subsection{Physics score test}\label{sec:score}

The omnibus channel is consistent against every alternative but spreads its power over 96 directions. Physically realisable deviations, however, move the outcome table along a small number of directions generated by perturbing device parameters. Let $\vartheta=(\vartheta_\psi,\,\vartheta_\eta)$ collect nine \emph{probe directions} and the nuisance:
\begin{equation}
\vartheta_\psi = \bigl(\mathrm{logit}\,e_Z,\ \mathrm{logit}\,e_X,\ \log p_{d0},\ \log p_{d1},\ \log\kappa,\ \log\rho,\ \Delta e_{\rm dec},\ \log c_{\rm dbl},\ \log\epsilon_X\bigr),\qquad \vartheta_\eta=\log\eta, \label{eq:directions}
\end{equation}
where the last three components vanish on honest links: $\Delta e_{\rm dec}$ lets the misalignment of decoy pulses differ from that of signal pulses, $c_{\rm dbl}$ rescales the double-click probability, and $\epsilon_X$ makes the receiver efficiency depend on Bob's basis. The score of window $t$ at the calibrated point is
\begin{equation}
u_t=\sum_{s,o}\bigl(n_{s,o}-E_{s,o}\bigr)\,\nabla_\vartheta\log p_{o\mid s}, \label{eq:score}
\end{equation}
and its Fisher information is $I_t=\sum_{s,o}E_{s,o}\,\nabla_\vartheta\log p_{o\mid s}\nabla_\vartheta\log p_{o\mid s}\tr$; the identity $\sum_o p_{o\mid s}\nabla\log p_{o\mid s}=0$ makes the multinomial correction term vanish (\cref{app:fisher}). Because $\eta$ is unknown, the scores along $\vartheta_\psi$ are replaced by their \emph{efficient} versions, the residuals of their projection onto the nuisance score,
\begin{equation}
u^{\rm eff}_t = u_{\psi,t} - I_{\psi\eta,t}I_{\eta\eta,t}^{-1}u_{\eta,t},\qquad
I^{\rm eff}_t = I_{\psi\psi,t} - I_{\psi\eta,t}I_{\eta\eta,t}^{-1}I_{\eta\psi,t}, \label{eq:effscore}
\end{equation}
which is Neyman's $C(\alpha)$ construction \cite{Neyman1959}: a small error in $\hat\eta_t$ then has no first-order effect on the statistic. Over a horizon of $M$ windows with $U_M=\sum_\tau(u^{\rm eff}_\tau-b_u)$ and $I_M=\sum_\tau I^{\rm eff}_\tau$, the score statistic is
\begin{equation}
S^{\rm ph}_M = \frac{1}{1+M/W}\;U_M\tr I_M^{-1/2}\,\Phi^{-1}\,I_M^{-1/2}U_M. \label{eq:scorestat}
\end{equation}
Under the honest hypothesis with Poissonian fluctuations, $\Phi=\mathbb{1}$ and $S^{\rm ph}_M$ is asymptotically $\chi^2_9$ \cite{Rao1948,LehmannRomano2005}; under a local alternative $\vartheta=\vartheta_0+h/\sqrt{N}$ it is noncentral $\chi^2_9$ with noncentrality $h\tr I^{\rm eff}h$, and no test of the same level has higher local power against alternatives in the span of the probe directions.

Real links fluctuate more than shot noise: window-to-window intensity jitter moves the decoy-ratio direction and slow polarisation and temperature drift move the error and background directions. We absorb this excess noise into a dispersion matrix estimated on commissioning windows,
\begin{equation}
\Phi = \mathcal{F}\Bigl[(1-\lambda_\Phi)\,\widehat{\mathrm{Cov}}(w)+\lambda_\Phi\,\mathrm{diag}\,\widehat{\mathrm{Cov}}(w)\Bigr],\qquad w_t=I_t^{-1/2}(u^{\rm eff}_t-b_u), \label{eq:disp}
\end{equation}
where $\lambda_\Phi=0.3$ is a shrinkage weight in the spirit of Ledoit and Wolf \cite{LedoitWolf2004} and $\mathcal{F}$ floors the eigenvalues at one, so that the honest model is never assumed to be quieter than shot noise. The same construction with the five free setting probabilities $(p_a,p_\alpha,p_\beta,p_\mathrm{d},p_\mathrm{v})$ defines the \emph{setting test} $S^{\rm set}_M$, which carries the signature of biased random-number generators.

\subsection{Drift-robust moving reference}\label{sec:moving}

All statistics above compare a window with commissioning. Slow honest drift then accumulates into the long-horizon statistics and raises their honest quantiles, which costs power. For each channel we therefore also compute a window-limited GLR that compares the recent block $(t-M,t]$ with the immediately preceding $R$ windows,
\begin{equation}
d_t = U_{(t-M,t]} - \frac{M}{R}\,U_{(t-M-R,\,t-M]},\qquad \mathrm{Var}(d_t)\simeq\Bigl(1+\frac{M}{R}\Bigr)\mathrm{Var}\bigl(U_{(t-M,t]}\bigr), \label{eq:moving}
\end{equation}
and uses $d_t$ in place of $U_M$ in \cref{eq:shift,eq:scorestat}. Static mismatch cancels exactly in $d_t$, and drift that is slow on the scale of $M+R$ windows contributes only through its change across the two blocks, whereas an abrupt change is fully visible during the first $M$ windows after it. An attack that ramps up more slowly than the reference length can evade the moving-reference channels; the fixed-reference channels remain and catch it eventually. We use $R=48$, fixed-reference horizons $M\in\{1,4,16\}$ and moving-reference horizons $M\in\{4,16,32\}$.

\subsection{Combination, thresholds and attribution}\label{sec:combine}

The monitor computes 16 channels per window: omnibus shift and energy, physics score and setting test at the fixed-reference horizons, and the moving-reference versions. Their honest distributions differ, because slow drift inflates long horizons, so each channel $c$ is standardised by its median $m_c$ and upper spread $q_c$ on attack-free \emph{training} links,
\begin{equation}
T_t=\max_c\ \frac{S_c(t)-m_c}{q_c-m_c},\qquad q_c=\text{0.999 quantile on training links}, \label{eq:combine}
\end{equation}
which equalises the contribution of each channel to the false-alarm rate. The alarm threshold is the $1-\alpha$ quantile of $T_t$ over attack-free \emph{validation} links, half of which contain benign loss faults, so that $\alpha$ is the per-window false-alarm rate. Training, validation and test links are disjoint.

At an alarm the monitor reports the physical constraint the anomaly violates. With the moving-reference sums over the last $M=16$ windows, the apparent displacement of the device parameters is $\hat\delta=I_M^{-1}d_t$ and its standardised value is $\hat\delta_j/\sqrt{(I_M^{-1})_{jj}(1+M/R)\Phi_{jj}}$; the parameter with the largest magnitude is reported. Attribution is defined over the nine device directions and does not apply to setting-test alarms. \Cref{alg:monitor} summarises one monitoring step.

\begin{algorithm}[t]
\caption{Monitoring step for window $t$}\label{alg:monitor}
\begin{algorithmic}[1]
\Require counts $n^{(t)}_{s,o}$; commissioning output of \cref{alg:calib}; channel standardisation $(m_c,q_c)$; threshold $\tau_\alpha$
\State $\hat\eta_t\gets\argmax_\eta\ell_t(\hat\psi,\eta)$ \Comment{Newton, \cref{eq:newton}}
\State $z_t\gets(r_t-b)/s$;\ \ $(u^{\rm eff}_t,I^{\rm eff}_t)\gets$ \cref{eq:effscore};\ \ setting scores likewise
\For{each channel $c$ (statistic, horizon $M$, fixed or moving reference)}
  \State update running sums; $S_c(t)\gets$ \cref{eq:shift}, energy, \cref{eq:scorestat} or its moving version \eqref{eq:moving}
\EndFor
\State $T_t\gets\max_c (S_c(t)-m_c)/(q_c-m_c)$
\If{$T_t>\tau_\alpha$}
  \State raise alarm; report $\argmax_j |\hat\delta_j|/\mathrm{se}(\hat\delta_j)$ \Comment{physical attribution}
\EndIf
\end{algorithmic}
\end{algorithm}

\section{Detectability limits}\label{sec:theory}

The honest-manifold formulation answers a question that a trained classifier cannot: how detectable is a given attack from telemetry, irrespective of the monitor?

\begin{definition}[Detectability exponent]\label{def:dstar}
Let $Q$ be the per-pulse distribution over settings and outcomes produced by an attack on a link with device parameters $\psi$, and $\Mhon_\psi$ the honest manifold \eqref{eq:manifold}. The detectability exponent of the attack is
\begin{equation}
\Dstar(Q)=\inf_{\eta\in(0,1]}\KL\bigl(Q\,\Vert\,P_{\psi,\eta}\bigr)\qquad\text{nats per pulse}. \label{eq:dstar}
\end{equation}
\end{definition}

$\Dstar$ is a property of the attack and the link alone. It measures the information per pulse that separates the attacked statistics from their closest honest explanation, where closeness is measured by the Kullback--Leibler divergence \cite{KullbackLeibler1951,CoverThomas2006}. For time-modulated attacks we average the divergence over one modulation period.

\begin{proposition}[Universal delay bound]\label{prop:lorden}
Consider windows of $N$ pulses that are i.i.d.\ given the link state, and let $\tau$ be the alarm time of any monitor whose average run length to false alarm is at least $\gamma$ under every honest state $P_{\psi,\eta}$, $\eta\in(0,1]$. Let $\eta^\ast$ attain the infimum in \eqref{eq:dstar}. If the link is honest at $\eta^\ast$ before the change, the worst-case mean detection delay satisfies
\begin{equation}
\sup_{\nu\ge1}\ \operatorname*{ess\,sup}\ \E_\nu\bigl[(\tau-\nu+1)^+\,\big\vert\,\mathcal{F}_{\nu-1}\bigr]\ \ge\ \bigl(1+o(1)\bigr)\,\frac{\log\gamma}{N\,\Dstar(Q)}\qquad(\gamma\to\infty), \label{eq:lorden}
\end{equation}
where $\nu$ is the change time and $\mathcal{F}_{\nu-1}$ the history before it.
\end{proposition}
\begin{proof}
A monitor that controls false alarms under every honest state controls them in particular under $P^{(N)}_{\psi,\eta^\ast}$, the law of one window at $\eta^\ast$. The change from $P^{(N)}_{\psi,\eta^\ast}$ to $Q^{(N)}$ is then a simple change between two known distributions whose per-window divergence is $\KL(Q^{(N)}\Vert P^{(N)}_{\psi,\eta^\ast})=N\,\Dstar(Q)$ by additivity of the divergence over independent pulses. Lorden's lower bound for simple changes \cite{Lorden1971}, in the general form of Lai \cite{Lai1998}, gives \eqref{eq:lorden}. Additional uncertainty about the pre-change state can only increase the delay.
\end{proof}

In the benchmark we control the per-window false-alarm probability $\alpha$ rather than the run length and plot the delay against $\log(1/\alpha)/(N\Dstar)$, which plays the role of the bound with $\gamma\approx1/\alpha$. The bound is asymptotic and concerns the worst-case mean delay; individual runs can fall below it. Two consequences make it useful in practice. First, it separates the performance of a monitor from the intrinsic visibility of an attack: a monitor whose delay tracks the bound cannot be improved by any other representation of the same telemetry, classical or quantum. Second, it exposes attacks that no monitor can see.

\begin{proposition}[Telemetry-blind attacks]\label{prop:blind}
$\Dstar(Q)=0$ if and only if $Q$ lies in the closure of $\Mhon_\psi$. In particular, a Trojan-horse probe whose back-reflection does not reach Bob ($q_{\rm bs}=0$ in \cref{sec:attacks}) has $\Dstar=0$, and for any monitor operating on the receiver telemetry of \cref{sec:settings} the probability of an alarm under this attack equals its false-alarm probability.
\end{proposition}
\begin{proof}
The divergence is non-negative and vanishes only for identical distributions; on the compact closure of $\{\eta\in(0,1]\}$ the infimum is attained, which gives the equivalence. The filtered Trojan-horse probe acts only on Alice's modulators: settings are drawn with the honest probabilities and Bob's outcome probabilities are those of $P_{\psi,\eta}$ at the true transmittance, so $Q=P_{\psi,\eta}$ exactly. The telemetry then has the same law with and without the attack, and any function of it, including any alarm rule, has the same law too.
\end{proof}

\Cref{prop:blind} is a statement about information, not about the shortcomings of a particular monitor. Attacks with $\Dstar=0$ must be excluded by device measures such as optical isolation and power monitoring \cite{Lucamarini2015,Ponosova2022}, or priced into the key rate. We verified in the benchmark that the numerically computed $\Dstar$ of the filtered probe is zero to machine precision.

\subsection{Local form and connection to the score test}\label{sec:local}

For weak attacks the divergence has a geometric form that connects \cref{def:dstar} to the score test of \cref{sec:score}.

\begin{proposition}[Local expansion]\label{prop:local}
Let $P_{\psi,\eta}$ be smooth in $\vartheta=(\vartheta_\psi,\log\eta)$ with Fisher information $J$ per pulse, and let the attack act locally along the device directions, $Q=P_{\vartheta_0+\epsilon h}$ with $h=(h_\psi,h_\eta)$. Then
\begin{equation}
\Dstar(Q)=\tfrac12\,\epsilon^2\,h_\psi\tr\bigl(J_{\psi\psi}-J_{\psi\eta}J_{\eta\eta}^{-1}J_{\eta\psi}\bigr)h_\psi+O(\epsilon^3). \label{eq:localdstar}
\end{equation}
\end{proposition}
\begin{proof}
For nearby members of a smooth family, $\KL(P_{\vartheta_0+\epsilon h}\Vert P_{\vartheta_0+\epsilon h'})=\tfrac12\epsilon^2(h-h')\tr J(h-h')+O(\epsilon^3)$. Minimising over the nuisance component $h'_\eta$ with $h'_\psi=0$ is a weighted least-squares problem whose solution $h'_\eta=h_\eta+J_{\eta\eta}^{-1}J_{\eta\psi}h_\psi$ leaves the Schur complement in \eqref{eq:localdstar}.
\end{proof}

The matrix in \cref{eq:localdstar} is the per-pulse efficient information \eqref{eq:effscore}. Hence, for attacks inside the span of the probe directions, the noncentrality of the score statistic after $m$ windows is $2mN\Dstar$, and the score test attains the delay bound up to the multiple-testing cost of the nine directions. Attacks outside the span still have $\Dstar>0$ if they leave the manifold; they are caught by the omnibus channel, at the price of more degrees of freedom.

\paragraph{A closed-form check.} For the RNG attack the outcome distribution is unchanged and the allocation of bit values is Bernoulli with parameter $\tfrac12+x$, $x=\delta\sin(2\pi t/T+\phi)$. Since $\KL(\mathrm{Bern}(\tfrac12+x)\Vert\mathrm{Bern}(\tfrac12))=2x^2+O(x^4)$ nats and the period average of $\sin^2$ is $1/2$,
\begin{equation}
\Dstar_{\rm RNG}=\delta^2+O(\delta^4). \label{eq:rngdstar}
\end{equation}
For $\delta=3\times10^{-4}$ the numerical evaluation of \cref{eq:dstar} on the full model gives $9.000\times10^{-8}$ nats per pulse, in agreement with $\delta^2$; with $N=10^7$ pulses per window and $\alpha=10^{-3}$ the delay bound is $\log(10^3)/0.9\approx7.7$ windows.

\paragraph{Interpretation for an adaptive adversary.} An eavesdropper who knows that \qtrace{} is in place wants to maximise her information subject to keeping $\Dstar$ small. \Cref{prop:lorden} turns this into a quantitative trade-off: she can remain undetected for about $\log(1/\alpha)/(N\Dstar)$ windows, during which she learns a number of key bits proportional to that time. The exponent is therefore the natural currency in which attack strength should be reported; \cref{sec:hw-info} measures the corresponding quantity, information leaked before the first alarm, on hardware.

\section{Benchmark design}\label{sec:benchmark}

\subsection{A twin with deliberate model mismatch}\label{sec:twin}

A detector built on a physical model and evaluated on data generated by the same model would be tested against itself. The twin used here therefore contains effects that the honest model of \cref{sec:family} does not describe, so that every result is obtained under model mismatch:
\begin{enumerate}[label=(\alph*)]
\item afterpulsing with probability up to 3\,\%, which the monitor knows only through a datasheet value carrying a $\pm10$\,\% error;
\item window-to-window jitter of each intensity class with relative standard deviation up to 1\,\%, independent between signal and decoy, which the honest model cannot represent with its single ratio $\rho$;
\item polarisation drift of the misalignments, an autoregressive process with per-window innovation up to $2\times10^{-4}$;
\item temperature-driven modulation of the dark-count probabilities with amplitude up to 10\,\%;
\item a random walk and a slow periodic modulation of the channel transmittance.
\end{enumerate}
Links are drawn with lengths of 5 to 60\,km at 0.18 to 0.22\,dB/km, receiver efficiencies of 0.10 to 0.50, misalignments of 0.5 to 2\,\%, dark-count probabilities of $2\times10^{-7}$ to $5\times10^{-6}$ per gate and detector efficiency ratios of 0.85 to 1.15; \cref{tab:twin} in \cref{app:params} lists all ranges. Each link is observed for 200 windows of $N=10^7$ pulses (one second at 10\,MHz), of which the first $W=60$ are attack-free commissioning windows. Counts are sampled from the multinomial laws \eqref{eq:mult}, so every cell, including vacuum double clicks, carries its own shot noise.

\subsection{Splits, attacks and faults}\label{sec:splits}

Each seed generates disjoint populations of links. The \emph{training} split (60 honest links, plus 30 links for each known attack used only by the supervised baseline) fixes the channel standardisation of \cref{eq:combine} and trains the learned baselines. The \emph{validation} split (50 honest links and 25 links with benign loss faults) fixes the thresholds of every method. The \emph{test} split contains 50 honest links, 25 links with loss faults (steps or ramps of 0.5 to 3\,dB), 15 links with a dark-count fault (a rise by a factor 1.5 to 5), 35 links for each of the six attack families and 20 for each of three compositions (PNS with blinding, time-shift with Trojan horse, intercept--resend with RNG bias). A \emph{long-link} split of 60 to 120\,km with receiver efficiencies of 0.05 to 0.20, a regime absent from training, contains 25 honest links and 15 links for each known attack.

Attacks start uniformly between 20 and 100 windows after commissioning, and windows are labelled positive only when an attack is active, so pre-onset windows of attacked links count as honest. Attack strengths are drawn log-uniformly over physically motivated ranges (\cref{tab:attackranges} in \cref{app:params}): intercepted fractions of 0.2 to 20\,\%, PNS and blinding fractions of 1 to 100\,\%, time-shift fractions of 2 to 100\,\% with mismatch factors of 0.1 to 0.5, back-scatter probabilities of $10^{-8}$ to $3\times10^{-6}$ per gate, and RNG bias amplitudes of $5\times10^{-5}$ to $2\times10^{-3}$ with periods of 10 to 80 windows. For every attacked run we also compute $\Dstar$ at the run's mean physical state, which allows each delay to be compared with its own bound.

\subsection{Comparison methods}\label{sec:baselines}

All comparison methods receive the same information as \qtrace{}: the 19 scalar telemetry quantities a deployed node exports (gains and QBERs per intensity and basis, vacuum yield, decoy-to-signal gain ratio, double-click rates, unsifted detector balance, click rate, bit and basis balance, $Y_1^L$, $e_1^U$ and the key rate), standardised per link against the same commissioning windows, together with their trailing 16-window means. We compare
\begin{itemize}
\item three \emph{industry monitors}: the static 11\,\% QBER threshold, a one-sided CUSUM on the standardised signal QBER \cite{Page1954}, and a two-sided CUSUM on the decoy-state estimates $Y_1^L$, $e_1^U$ and the gain ratio, i.e.\ the decoy consistency check of deployed systems made sequential;
\item four \emph{unsupervised detectors} trained on honest data only: an isolation forest \cite{Liu2008IF}, a one-class SVM \cite{Scholkopf2001}, a multilayer-perceptron autoencoder, and a one-class SVM on a fidelity quantum kernel \cite{Havlicek2019,SchuldKilloran2019} with a six-qubit ZZ feature map on principal components, evaluated with exact statevectors that we verified against Qiskit \cite{Qiskit2024} to unit fidelity;
\item a \emph{supervised} gradient-boosted classifier (XGBoost \cite{ChenGuestrin2016}) trained on the three known attack families.
\end{itemize}
Standard implementations from scikit-learn \cite{Pedregosa2011} and XGBoost are used with fixed hyperparameters listed in the repository. Every method's threshold is the $1-\alpha$ quantile of its score on the same validation links, so all methods operate at the same per-window false-alarm rate.

\subsection{Metrics and protocol}\label{sec:metrics}

The primary metric is the fraction of attacked links on which a method raises an alarm after onset, at a per-window false-alarm rate of $\alpha=10^{-3}$ ($10^{-2}$ in \cref{app:results}). We also report the window-level AUROC (attack-active windows against all windows of honest test links), the detection delay in windows, the key generated between onset and first alarm (key exposed), and the fraction of honest, loss-fault and long-link honest runs with at least one alarm. All design decisions, including the horizons, the reference length and the shrinkage constants, were fixed on development seeds 0 to 3. The reported results use seeds \SeedList{}, which played no role in development; means are reported with 95\,\% Student-$t$ confidence intervals over seeds. In total the evaluation comprises \NRunsTotal{} links and \NWindowsTotal{} telemetry windows; one seed takes about \WallPerSeed{} minutes on a single CPU core.

\section{Simulation results}\label{sec:results}

\subsection{Detection at a matched false-alarm rate}\label{sec:res-main}

\begin{figure}[!tbp]
\centering
\includegraphics[width=\textwidth]{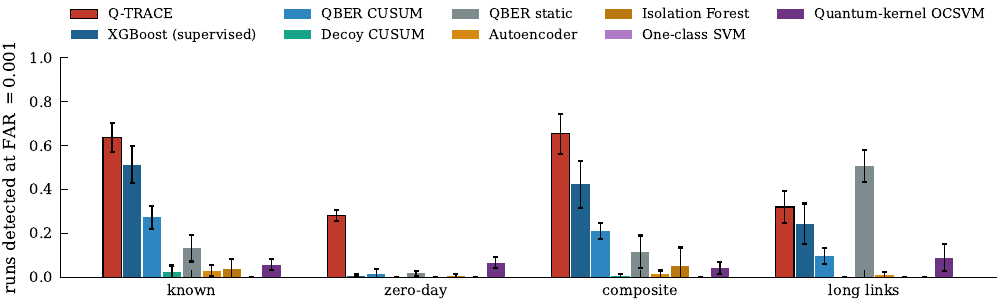}
\caption{Fraction of attacked links detected at a per-window false-alarm rate of $10^{-3}$, by attack group. Error bars are 95\,\% confidence intervals over held-out seeds.}
\label{fig:summary}
\end{figure}

\Cref{tab:detection} and \cref{fig:summary} give the headline comparison. At a per-window false-alarm rate of $10^{-3}$, \qtrace{} raised an alarm on \DetKnownQT\,\% of links under the known attack families, \DetZeroDayQT\,\% under the zero-day families and \DetCompQT\,\% under composite attacks. The strongest comparison method in each group reached \DetKnownBest\,\% (\DetKnownBestName), \DetZeroDayBest\,\% (\DetZeroDayBestName) and \DetCompBest\,\% (\DetCompBestName). The zero-day figure of the quantum-kernel detector should not be read as detection: the same method alarms on \FAHonestQKoneclassSVM\,\% of attack-free links, more often than it alarms on zero-day attacks. The supervised classifier, which was trained on exactly the known families, is the closest competitor on known attacks and falls to \DetZeroDayXGBoostsup\,\% on zero-day families, which is the behaviour one expects from a model that has learned what particular attacks look like rather than what honest links look like. The decoy-state CUSUM, which encodes the consistency check most deployed systems perform, detects only \DetKnownDecoyCUSUM\,\% of known attacks: the decoy-state estimates are noisy functionals of the gains, and stealthy blinding and weak PNS move them by less than their window-to-window fluctuation. The window-level AUROC (\cref{tab:auroc}) gives a similar picture, with \qtrace{} at \AucKnownQT, \AucZeroDayQT{} and \AucCompQT{} for the three groups, with one exception: on long links the supervised classifier reaches \AucShiftXGB{} against \AucShiftQT{} for \qtrace{}. AUROC weights all thresholds equally, whereas operation happens at a low false-alarm rate, where the detection rates of \cref{tab:detection} are the relevant measure.

\begin{table}[!tbp]
\centering
\caption{Percentage of attacked links detected, by attack group, and percentage of attack-free test links with at least one alarm (FA), at a per-window false-alarm rate of $10^{-3}$ set on validation links. ``Long links'' is the 60 to 120\,km regime absent from training. Mean\,$\pm$\,95\,\% confidence interval over \NSeeds{} held-out seeds.}
\label{tab:detection}
\small
\resizebox{\textwidth}{!}{\begin{tabular}{lccccccc}
\toprule
Method & Known & Zero-day & Composite & Long links & FA honest & FA loss & FA long \\
\midrule
\textbf{Q-TRACE} & \textbf{63.6\,$\pm$\,6.6} & \textbf{28.0\,$\pm$\,2.5} & \textbf{65.3\,$\pm$\,9.1} & \textbf{32.0\,$\pm$\,7.5} & \textbf{5.2\,$\pm$\,3.8} & \textbf{0.0\,$\pm$\,0.0} & \textbf{3.2\,$\pm$\,2.2} \\
XGBoost (sup.) & 51.2\,$\pm$\,8.5 & 0.6\,$\pm$\,0.6 & 42.3\,$\pm$\,10.8 & 24.4\,$\pm$\,9.2 & 0.8\,$\pm$\,1.4 & 8.8\,$\pm$\,6.5 & 0.0\,$\pm$\,0.0 \\
QBER CUSUM & 27.2\,$\pm$\,5.3 & 1.3\,$\pm$\,2.5 & 21.0\,$\pm$\,3.8 & 9.8\,$\pm$\,3.7 & 1.2\,$\pm$\,2.2 & 1.6\,$\pm$\,2.7 & 0.0\,$\pm$\,0.0 \\
Decoy CUSUM & 2.5\,$\pm$\,2.8 & 0.0\,$\pm$\,0.0 & 0.3\,$\pm$\,0.9 & 0.0\,$\pm$\,0.0 & 0.0\,$\pm$\,0.0 & 3.2\,$\pm$\,2.2 & 0.0\,$\pm$\,0.0 \\
QBER static & 13.1\,$\pm$\,6.0 & 1.7\,$\pm$\,1.3 & 11.7\,$\pm$\,7.3 & 50.7\,$\pm$\,7.4 & 2.8\,$\pm$\,2.8 & 0.8\,$\pm$\,2.2 & 45.6\,$\pm$\,16.7 \\
Autoencoder & 3.0\,$\pm$\,2.7 & 0.0\,$\pm$\,0.0 & 1.3\,$\pm$\,1.7 & 0.9\,$\pm$\,1.5 & 0.0\,$\pm$\,0.0 & 4.0\,$\pm$\,8.6 & 0.0\,$\pm$\,0.0 \\
Isolation forest & 3.6\,$\pm$\,4.6 & 0.4\,$\pm$\,1.1 & 5.0\,$\pm$\,8.5 & 0.0\,$\pm$\,0.0 & 0.0\,$\pm$\,0.0 & 12.0\,$\pm$\,8.6 & 0.0\,$\pm$\,0.0 \\
One-class SVM & 0.0\,$\pm$\,0.0 & 0.0\,$\pm$\,0.0 & 0.0\,$\pm$\,0.0 & 0.0\,$\pm$\,0.0 & 0.0\,$\pm$\,0.0 & 0.0\,$\pm$\,0.0 & 0.0\,$\pm$\,0.0 \\
QK one-class SVM & 5.7\,$\pm$\,2.5 & 6.7\,$\pm$\,2.4 & 4.3\,$\pm$\,2.8 & 8.9\,$\pm$\,6.2 & 10.4\,$\pm$\,3.2 & 12.8\,$\pm$\,4.2 & 7.2\,$\pm$\,5.4 \\
\bottomrule
\end{tabular}
}
\end{table}

\begin{figure}[!tbp]
\centering
\includegraphics[width=\textwidth]{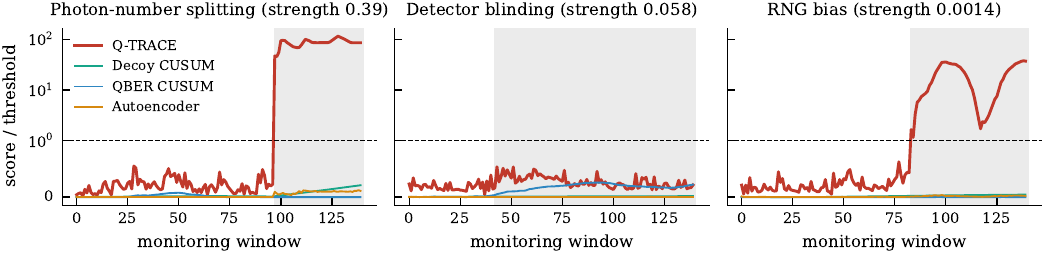}
\caption{Alarm statistics divided by their thresholds on three test links (shaded: attack active). Values above one are alarms. The vertical axis is linear below one and logarithmic above.}
\label{fig:traces}
\end{figure}

\Cref{fig:traces} shows what these numbers mean on individual links. For a PNS attack that leaves the gain and QBER unchanged, the \qtrace{} statistic crosses its threshold within a few windows of onset, while the QBER and decoy-state CUSUMs, scaled to their own thresholds, stay below one. For stealthy blinding, which reproduces gain and QBER by construction, the alarm is driven by the vanishing double clicks and vacuum clicks. For the modulated RNG attack the energy and setting channels respond, whereas monitors built on time-averaged balance do not.

\subsection{Attack families and strength}\label{sec:res-family}

\begin{figure}[!tbp]
\centering
\includegraphics[width=\textwidth]{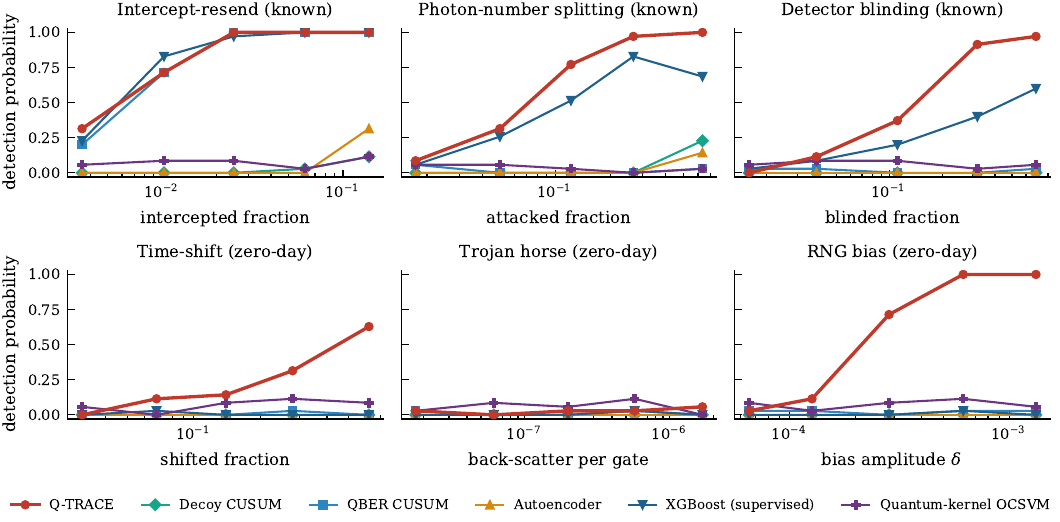}
\caption{Detection probability at a per-window false-alarm rate of $10^{-3}$ against attack strength, in quintile bins of strength. Pooled over held-out seeds.}
\label{fig:strength}
\end{figure}

\Cref{fig:strength} resolves detection by family and attack strength, and \cref{tab:perfamily} lists the per-family detection rates of all methods. Intercept--resend is the one family for which a QBER monitor is close to optimal, because the attack does nothing except raise the QBER; the QBER CUSUM detects \FamInterceptresendQC\,\% and \qtrace{} \FamInterceptresendQT\,\%. For PNS (\FamPNSQT\,\% against \FamPNSXGB\,\% for the supervised classifier and \FamPNSQC\,\% for the QBER CUSUM) and blinding (\FamBlindingQT\,\% against \FamBlindingXGB\,\%) the advantage comes from constraints that the QBER does not see. Among the zero-day families, \qtrace{} detects \FamRNGbiasQT\,\% of RNG-bias attacks, for which every comparison method stays near its false-alarm level, and \FamTimeshiftQT\,\% of time-shift attacks. Detection of Trojan-horse probes with back-scatter is low for every method (\qtrace{} \FamTrojanhorseQT\,\%): the back-scattered background is small compared with the afterpulse background of most links, and the theory below shows that the information available is correspondingly small.

\begin{table}[!htb]
\centering
\caption{Percentage of attacked test links detected, per attack family, at a per-window false-alarm rate of $10^{-3}$ (mean over seeds; best method per row in bold).}
\label{tab:perfamily}
\small
\begin{tabular}{lccccccccc}
\toprule
Family & \rotatebox{70}{Q-TRACE} & \rotatebox{70}{XGBoost (sup.)} & \rotatebox{70}{QBER CUSUM} & \rotatebox{70}{Decoy CUSUM} & \rotatebox{70}{QBER static} & \rotatebox{70}{Autoencoder} & \rotatebox{70}{Isolation forest} & \rotatebox{70}{One-class SVM} & \rotatebox{70}{QK one-class SVM} \\
\midrule
Intercept-resend & \textbf{80.6} & 80.6 & 78.3 & 2.9 & 36.6 & 6.3 & 9.1 & 0.0 & 7.4 \\
PNS & \textbf{62.9} & 46.9 & 1.7 & 4.6 & 1.7 & 2.9 & 1.7 & 0.0 & 3.4 \\
Blinding & \textbf{47.4} & 26.3 & 1.7 & 0.0 & 1.1 & 0.0 & 0.0 & 0.0 & 6.3 \\
Time-shift & \textbf{24.0} & 0.6 & 0.6 & 0.0 & 2.9 & 0.0 & 0.6 & 0.0 & 6.9 \\
Trojan horse & 2.9 & 0.6 & 1.1 & 0.0 & 1.1 & 0.0 & 0.6 & 0.0 & \textbf{5.7} \\
RNG bias & \textbf{57.1} & 0.6 & 2.3 & 0.0 & 1.1 & 0.0 & 0.0 & 0.0 & 7.4 \\
PNS + blinding & \textbf{81.0} & 57.0 & 1.0 & 1.0 & 2.0 & 2.0 & 6.0 & 0.0 & 3.0 \\
Time-shift + Trojan & \textbf{29.0} & 3.0 & 0.0 & 0.0 & 1.0 & 0.0 & 0.0 & 0.0 & 6.0 \\
IR + RNG bias & \textbf{86.0} & 67.0 & 62.0 & 0.0 & 32.0 & 2.0 & 9.0 & 0.0 & 4.0 \\
\bottomrule
\end{tabular}

\end{table}

\subsection{Delay against the information-theoretic bound}\label{sec:res-bound}

\begin{figure}[!htb]
\centering
\includegraphics[width=0.56\textwidth]{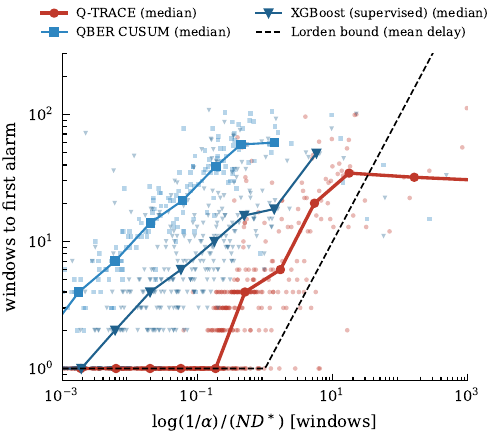}
\caption{Windows of post-onset data needed for the first alarm against the information-theoretic bound of \cref{prop:lorden} for each detected attack (points) and binned medians (lines). The dashed line is $\max(1,\text{bound})$. The bound constrains the mean delay, so individual runs can fall below it.}
\label{fig:bound}
\end{figure}

\Cref{fig:bound} places every detected attack on the plane of its own delay bound $\log(1/\alpha)/(N\Dstar)$ and its observed delay. The median delay of \qtrace{} lies on the bound for attacks with bounds below one window and stays within a small factor of it up to bounds of about ten windows. The QBER CUSUM and the supervised classifier detect fewer attacks, and those they detect require between one and two orders of magnitude more data than the bound. \Cref{tab:bound} quantifies the same effect as detection rates: for attacks whose bound is below one window, \qtrace{} detects \BoundLtOneQT\,\% (the supervised classifier \BoundLtOneXGB\,\%, the QBER CUSUM \BoundLtOneQC\,\%); for bounds between one and ten windows the rates are \BoundOneTenQT\,\%, \BoundOneTenXGB\,\% and \BoundOneTenQC\,\%.

\begin{table}[!tbp]
\centering
\caption{Percentage of attacked links detected, grouped by the delay bound $\log(1/\alpha)/(N\Dstar)$ of each run in windows, at $\alpha=10^{-3}$. Runs are pooled over seeds; the last row gives the number of attack runs per bin.}
\label{tab:bound}
\small
\begin{tabular}{lcccc}
\toprule
Method & bound $<1$ & bound 1--10 & bound 10--100 & bound $>100$ \\
\midrule
\textbf{Q-TRACE} & \textbf{93.7} & \textbf{25.4} & \textbf{7.8} & \textbf{2.6} \\
XGBoost (sup.) & 62.0 & 6.4 & 1.8 & 0.3 \\
QBER CUSUM & 31.7 & 3.4 & 1.8 & 1.0 \\
Decoy CUSUM & 2.6 & 0.0 & 0.0 & 0.0 \\
QBER static & 19.9 & 13.6 & 11.5 & 10.9 \\
Autoencoder & 3.7 & 0.0 & 0.0 & 0.0 \\
Isolation forest & 4.1 & 0.0 & 0.0 & 0.3 \\
One-class SVM & 0.0 & 0.0 & 0.0 & 0.0 \\
QK one-class SVM & 6.7 & 6.1 & 4.6 & 8.6 \\
\midrule
Attack runs & 492 & 264 & 217 & 302 \\
\bottomrule
\end{tabular}

\end{table}

The gap between \qtrace{} and the bound for bounds between one and a hundred windows has a clear origin. The bound assumes a stationary honest link, whereas the twin drifts. In a development run on seed 1 with all nuisance drift switched off, the first version of the monitor detected 74\,\% of attacked links and 93\,\% of those with bounds below 20 windows; with drift switched on the same monitor detected 39\,\%, and the moving-reference channels of \cref{sec:moving} raised this to 59\,\% before the configuration was frozen. Slow honest drift, not the statistical efficiency of the monitor, is therefore the binding limit in the realistic regime. An extension of the bound to drifting pre-change laws is left for future work.

\subsection{False alarms, benign faults and unseen regimes}\label{sec:res-fa}

\begin{figure}[!tbp]
\centering
\includegraphics[width=\textwidth]{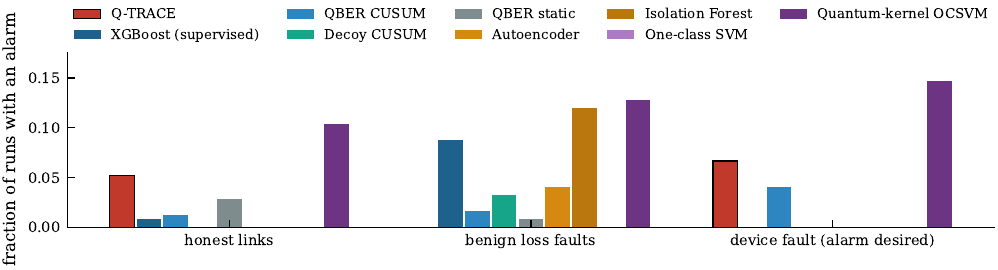}
\caption{Fraction of test links with at least one alarm at a per-window false-alarm rate of $10^{-3}$: attack-free links, links with benign loss faults (steps or ramps of 0.5 to 3\,dB) and links with a dark-count fault, where an alarm is desirable.}
\label{fig:fa}
\end{figure}

\Cref{fig:fa} and the last three columns of \cref{tab:detection} show how often each method alarms on attack-free links. Over the 140 monitoring windows of a link the per-window rate of $10^{-3}$ corresponds to a run-level rate of roughly 13\,\% for independent windows; on nominal honest links all methods stay below this, \qtrace{} at \FAHonestQT\,\%. \qtrace{} raised an alarm on \FALossQT\,\% of links with benign loss faults, because a loss step changes only the nuisance parameter; the isolation forest (\FALossIsolationforest\,\%), the quantum-kernel one-class SVM (\FALossQKoneclassSVM\,\%) and the autoencoder (\FALossAutoencoder\,\%) treat the same steps as anomalies. On honest long links, the static QBER threshold alarms on \FALongQBERstatic\,\% of runs because the honest QBER of long links can exceed 11\,\%; its apparently high detection rate in the long-link regime (\DetShiftQBERstatic\,\%) is an artefact of this and not evidence of detection. \qtrace{} alarms on \FALongQT\,\% of honest long links and detects \DetShiftQT\,\% of attacks there, lower than in the nominal regime because fewer clicks per window carry less information. Dark-count faults are detected by none of the methods with high probability: with afterpulse backgrounds of order $10^{-4}$ per gate, a rise of the intrinsic dark-count probability from $10^{-6}$ by a factor of a few changes the total background only marginally.

\subsection{Physical attribution of alarms}\label{sec:res-attr}

\begin{figure}[!htb]
\centering
\includegraphics[width=0.78\textwidth]{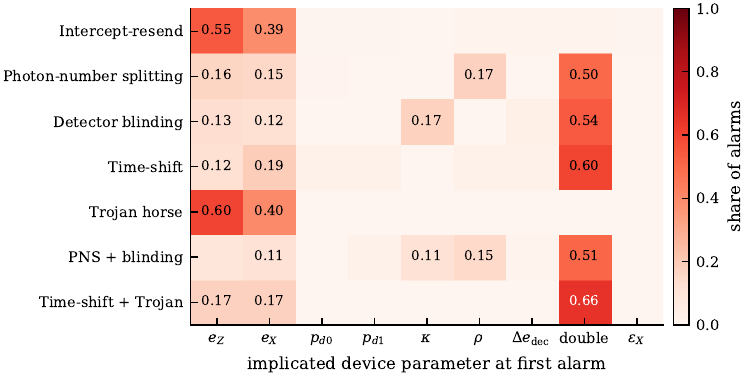}
\caption{Device parameter implicated at the first alarm of \qtrace{}, by attack family (share of detected runs; entries below 0.1 not printed). RNG bias is excluded because attribution is defined over device directions only.}
\label{fig:attr}
\end{figure}

Because the score channel is built from physical directions, an alarm comes with a statement of which constraint was violated. \Cref{fig:attr} shows, for each attack family, the device parameter with the largest standardised displacement at the first alarm. Blinding and PNS alarms are most often attributed to the double-click direction (\AttrBlindingShare\,\% and \AttrPNSShare\,\% respectively), consistent with blinded detectors that never double-click and with the altered multi-photon statistics of a lossless PNS channel. Intercept--resend alarms are most often attributed to an optical-error direction (\AttrInterceptresendShare\,\%), as \cref{eq:ir} predicts. The attribution is a diagnostic aid, not a classifier: it was not tuned on attack data, and it does not apply to the RNG attack, whose alarms come from the setting channel.

\subsection{The quantum-kernel baseline}\label{sec:res-qk}

The one-class SVM on a quantum fidelity kernel performs close to chance (AUROC \AucKnownQK{} on known attacks) and has the highest false-alarm rate on loss faults. We do not read this as a statement about quantum machine learning in general. The kernel acts on a six-dimensional projection of standardised telemetry scalars, and \cref{prop:lorden} implies that no representation of those scalars, quantum or classical, can extract information that the scalars do not contain; much of the relevant information sits in cells, such as vacuum double clicks, that the scalars summarise poorly. The result does suggest that, for this problem, the choice of statistic matters more than the choice of model class.

\section{A hardware study on an IBM Heron processor}\label{sec:hardware}

The simulation study tests \qtrace{} against a twin whose honest noise we wrote ourselves. A hardware experiment tests the parts of the method that do not depend on optical physics, namely calibration against a device's own commissioning data, false-alarm control under real noise and drift, the value of tests that use the full outcome table, and agreement with information-theoretic predictions, on a device whose noise is not generated by any model of ours. It also makes the eavesdropper physical: her interaction is a unitary acting on the transmitted qubit and her own ancilla, and her information can be measured.

\subsection{Multiplexed BB84 links and physical eavesdroppers}\label{sec:hw-design}

\begin{figure}[!tbp]
\centering
\begin{quantikz}[row sep=0.35cm, column sep=0.32cm, font=\small]
\lstick{data: $\ket{0}$} & \gate{X^{a}} & \gate{H^{\alpha}} & \slice{} & \gate[2]{U_{\rm Eve}} & \slice{} & \gate{H^{\beta}} & \meter{} & \setwiretype{c} \rstick{Bob}\\
\lstick{ancilla: $\ket{0}$} & & & & & & \gate{(HS^\dagger)^{y}} & \meter{} & \setwiretype{c} \rstick{Eve}
\end{quantikz}
\caption{One BB84 link. Alice prepares $\ket{a}$ in basis $\alpha$; Eve applies $U_{\rm Eve}\in\{\mathbb{1},\,R_{ZX}(\theta),\,R_{XX}(\theta),\,U_{\rm AD}(\gamma)\}$; Bob measures in basis $\beta$. Eve's ancilla is read in the $Y$ basis ($y=1$) for the entangling probes and in the $Z$ basis for the damping probe. Barriers (dashed) align all links of a circuit.}
\label{fig:circuit}
\end{figure}
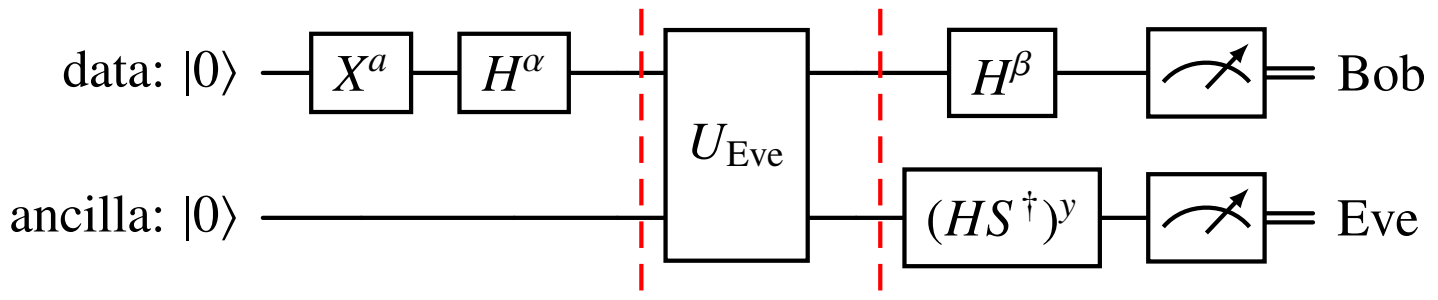

Each BB84 link uses two neighbouring qubits: a data qubit that carries Alice's state to Bob's measurement and an ancilla owned by Eve. Pairs are ranked by two-qubit-gate error plus both readout errors, subject to cut-offs of 2\,\% and 5\,\% respectively, and are chosen mutually non-adjacent so that no coupler connects qubits of different links. On \HwBackend{} this gave \HwLinks{} links; \cref{tab:hwcal} in \cref{app:hardware} summarises their calibration. One circuit realises one BB84 round per link and per shot (\cref{fig:circuit}). The eight settings $(a,\alpha,\beta)$ rotate in a Latin-square pattern, link $i$ in circuit $c$ using setting $(i+c)\bmod 8$, so that every circuit contains every setting and every link sees every setting once per pass of eight circuits.

The experiment comprises two commissioning passes, in which every link is honest, followed by six monitoring epochs of one pass each. In the monitoring epochs every link is honest in two epochs and attacked in four, with four distinct probes drawn from a balanced, seeded schedule. Ten probes are used.

\paragraph{Entangling probe.} Eve couples her ancilla to the data qubit with $R_{ZX}(\theta)=\exp(-i\theta Z\otimes X/2)$, which learns the $Z$-basis bit, or with $R_{XX}(\theta)=\exp(-i\theta X\otimes X/2)$, which learns the $X$-basis bit. For a data qubit in an eigenstate of the coupled Pauli operator with eigenvalue $(-1)^a$, the ancilla evolves to
\begin{equation}
\ket{e_\pm}=\cos\tfrac{\theta}{2}\ket{0}\mp i\sin\tfrac{\theta}{2}\ket{1},\qquad \braket{e_+}{e_-}=\cos\theta. \label{eq:probestates}
\end{equation}
A data qubit prepared in the conjugate basis becomes entangled with the ancilla, and its coherence is multiplied by $\braket{e_+}{e_-}$, so that Bob's error probability in the conjugate basis is
\begin{equation}
D(\theta)=\frac{1-\cos\theta}{2}=\sin^2\frac{\theta}{2}. \label{eq:disturb}
\end{equation}
Eve's optimal guess of the bit in the coupled basis is the Helstrom measurement \cite{Helstrom1976} discriminating $\ket{e_+}$ from $\ket{e_-}$, whose error probability is
\begin{equation}
P_{\rm err}(\theta)=\tfrac12\Bigl(1-\sqrt{1-\lvert\braket{e_+}{e_-}\rvert^2}\Bigr)=\frac{1-\sin\theta}{2}, \label{eq:helstrom}
\end{equation}
attained by a $Y$-basis measurement. Her information per bit in the coupled basis is $I(\theta)=1-\h\bigl(P_{\rm err}(\theta)\bigr)$. Eliminating $\theta$ with $\sin\theta=2\sqrt{D(1-D)}$ gives
\begin{equation}
I(D)=1-\h\!\Bigl(\tfrac12-\sqrt{D(1-D)}\Bigr), \label{eq:infodisturb}
\end{equation}
which has the same functional form as the optimal information--disturbance trade-off for individual attacks on BB84 derived by Fuchs \emph{et al.} \cite{Fuchs1997}. At $\theta=\pi/2$ the ancilla states are orthogonal and the probe is an intercept--resend attack in the coupled basis. Eve's choice between $R_{ZX}$ and $R_{XX}$ is balanced so that, in every pass, she matches Alice's basis on exactly half of the sifted settings; the average sifted QBER increase is then $\Delta Q_{\rm ideal}=\tfrac12\sin^2(\theta/2)$. A purely random choice per circuit would leave a probability of $1/16$ that an epoch is never disturbed, which a noisy rehearsal of the experiment revealed and which the balanced schedule removes.

\paragraph{Amplitude-damping probe.} The coupling $U_{\rm AD}(\gamma)=\mathrm{CX}_{e\to d}\,\mathrm{CR}_Y(2\arcsin\sqrt\gamma)_{d\to e}$ transfers the excitation of the data qubit to the ancilla with probability $\gamma$. It produces bit-asymmetric errors, $\gamma$ for $a=1$ and none for $a=0$ in the $Z$ basis and $(1-\sqrt{1-\gamma})/2$ in the $X$ basis, the qubit analogue of an efficiency-mismatch attack. Eve's ancilla is $\ket{1}$ only after a decay, so her information about $Z$-basis bits is $\h(\gamma/2)-\tfrac12\h(\gamma)$.

\paragraph{Execution.} All circuits were transpiled with a fixed layout and no routing, executed with the Qiskit Runtime sampler \cite{Qiskit2024} without error mitigation, twirling or dynamical decoupling (the monitor must work on raw device statistics), and with per-shot readout so that detection could later be studied as a function of the number of rounds. A pilot job of eight circuits with 400 shots measured the cost per shot (\HwPilotSec\,s); the main job of 64 circuits with \HwShots{} shots each used \HwMainSec\,s, a total of \HwTotalSec\,s of processor time for \HwRounds{} million link-rounds. Before any processor time was spent, the full pipeline was validated on a noisy simulation of the device.

\subsection{Tests}\label{sec:hw-tests}

For every link and monitoring epoch three tests compare the epoch with a reference. \emph{\qtrace{}} transfers the structure of the simulation monitor to the qubit setting: an error channel (two-sided change of the $Z$- and $X$-basis error rates, $\chi^2_2$), an asymmetry channel (change of $P(1\mid a=0)-P(0\mid a=1)$ per basis, $\chi^2_2$) and an omnibus channel (two-sample likelihood-ratio test over all eight settings, $\chi^2_8$), combined with a Bonferroni correction. The \emph{pooled QBER} test is the one-sided two-proportion test on the sifted QBER that deployed systems use; the \emph{per-basis QBER} test applies one-sided tests to both bases with a Bonferroni correction. Dispersion factors for between-circuit fluctuations were estimated from the two commissioning passes and were close to one ($\phi=\HwPhi$ for the omnibus channel, $\phi=\HwPhiQ$ for the pooled QBER), confirming that circuits repeat at the shot-noise level.

\subsection{Pre-registered analysis: the reference matters}\label{sec:hw-pre}

\begin{figure}[!tbp]
\centering
\includegraphics[width=\textwidth]{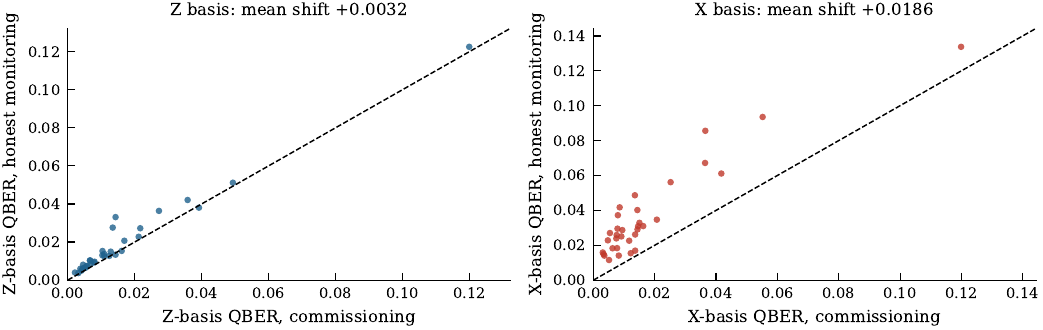}
\caption{Honest links on \HwBackend{}: QBER in the commissioning passes against QBER in the honest monitoring epochs, per basis. The $X$ basis shifts by \HwShiftX\,\% on every link, the $Z$ basis by \HwShiftZ\,\%: commissioning circuits did not reproduce the operating conditions of monitoring circuits.}
\label{fig:shift}
\end{figure}

The analysis plan fixed before the run used the two commissioning passes as reference. With this reference every test failed: \qtrace{}, the per-basis QBER test and the pooled QBER test flagged \HwPreFAQT\,\%, \HwPreFAQB\,\% and \HwPreFAQP\,\% of the \HwHonestN{} honest link-epochs at $\alpha=10^{-3}$ (\cref{tab:hwpre} in \cref{app:hardware}). The cause is physical and is shown in \cref{fig:shift}. Relative to commissioning, the honest monitoring epochs show a QBER increase of \HwShiftX\,\% in the $X$ basis, on all \HwShiftXPos{} links, and of only \HwShiftZ\,\% in the $Z$ basis. The two commissioning passes agree to $\HwPassDiff$, and the increase shows no trend over the six epochs, so slow drift is excluded. Commissioning circuits contain no probe gates, so their qubits are measured immediately after preparation; in monitoring circuits the barriers make every data qubit, including those of honest links, wait for the longest probe on the chip, while two-qubit gates act on other links. A dephasing-dominated error that appears only under these conditions is what one expects from the added idle time together with crosstalk from concurrent coupler activation. The present data cannot separate the two contributions.

The monitors were right. The operating conditions had changed, and a reference acquired under different conditions describes a different device. The same failure would occur in an optical system calibrated without traffic, with different detector gating, or at a different temperature. The practical lesson is that the commissioning data of any telemetry monitor, QBER-based or otherwise, must be acquired under the operating conditions of the monitored traffic; in the hardware setting this means inserting timing-matched delays and the same concurrent gate activity into the commissioning circuits.

\subsection{Condition-matched analysis}\label{sec:hw-matched}

Each link was honest in two monitoring epochs that ran under exactly the conditions of the attacked epochs. Using these epochs as reference, we test each attacked epoch against the link's two honest epochs and estimate the false-alarm rate by testing each honest epoch against the link's other honest epoch. This analysis was defined after the pre-registered one had failed and is reported as such.

\begin{figure}[!tbp]
\centering
\includegraphics[width=\textwidth]{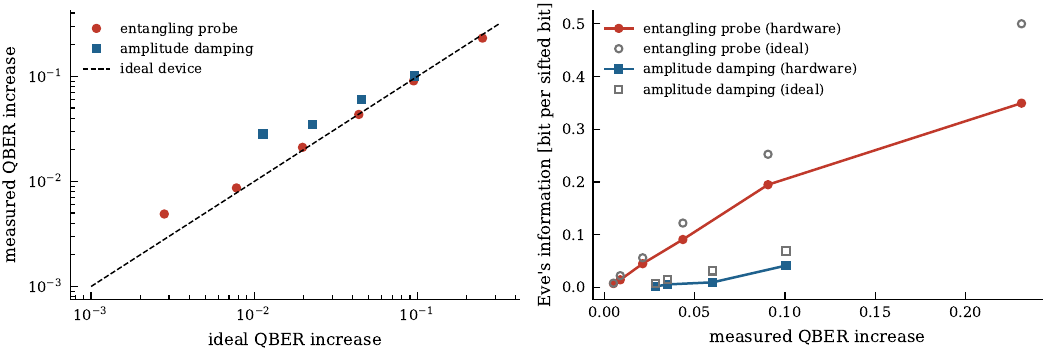}
\caption{Probe physics on \HwBackend{} (condition-matched reference). (a) Measured against ideal increase of the sifted QBER; the dashed line is the ideal device. (b) Eve's information per sifted bit, measured on her ancilla, against the measured QBER increase; open symbols are the ideal-device values of \cref{eq:infodisturb} and of the damping channel.}
\label{fig:hwphys}
\end{figure}

With the condition-matched reference none of the \HwHonestN{} honest link-epochs was flagged by any test (\cref{tab:hwmatched}). The honest $p$-values are, if anything, conservative: 3\,\% fall below 0.05. The measured QBER increase of every entangling probe agrees with $\Delta Q_{\rm ideal}$ across two decades (\cref{fig:hwphys}a); for example $\theta=0.60$ gives \HwThetaDMeas\,\% against \HwThetaDIdeal\,\% predicted and $\theta=0.90$ gives \HwThetaEMeas\,\% against \HwThetaEIdeal\,\%. The damping probe raises the QBER above its ideal value at small $\gamma$, because its three two-qubit gates add their own error, an honest feature of a physical eavesdropper whose apparatus is imperfect.

\begin{table}[!tbp]
\centering
\caption{Hardware detection with the condition-matched reference: fraction of link-epochs flagged at $\alpha=10^{-3}$; $n$ is the number of link-epochs, $\Delta Q_{\rm ideal}$ the ideal and $\Delta Q_{\rm meas}$ the measured increase of the sifted QBER.}
\label{tab:hwmatched}
\small
\begin{tabular}{lccccccc}
\toprule
Probe & $n$ & $\Delta Q_{\rm ideal}$ (\%) & $\Delta Q_{\rm meas}$ (\%) & Q-TRACE & QBER/basis & QBER pooled & Omnibus \\
\midrule
Honest (no probe) & 66 & 0.00 & 0.00 & 0.00 & 0.00 & 0.00 & 0.00 \\
\midrule
$\theta=0.15$ & 14 & 0.28 & 0.49 & 1.00 & 0.93 & 0.64 & 1.00 \\
$\theta=0.25$ & 13 & 0.78 & 0.87 & 0.92 & 0.85 & 0.85 & 0.92 \\
$\theta=0.40$ & 14 & 1.97 & 2.11 & 1.00 & 1.00 & 1.00 & 1.00 \\
$\theta=0.60$ & 14 & 4.37 & 4.34 & 1.00 & 1.00 & 1.00 & 1.00 \\
$\theta=0.90$ & 13 & 9.46 & 9.05 & 1.00 & 1.00 & 1.00 & 1.00 \\
$\theta=1.57$ & 13 & 25.00 & 23.10 & 1.00 & 1.00 & 1.00 & 1.00 \\
\midrule
$\gamma=0.03$ & 13 & 1.13 & 2.82 & 1.00 & 1.00 & 1.00 & 1.00 \\
$\gamma=0.06$ & 12 & 2.26 & 3.49 & 1.00 & 1.00 & 1.00 & 1.00 \\
$\gamma=0.12$ & 12 & 4.55 & 5.99 & 1.00 & 1.00 & 1.00 & 1.00 \\
$\gamma=0.25$ & 14 & 9.60 & 10.05 & 1.00 & 1.00 & 1.00 & 1.00 \\
\bottomrule
\end{tabular}

\end{table}

All probes with $\Delta Q_{\rm ideal}\ge 2$\,\% are detected on every link-epoch by every test. The informative case is the weakest probe, $\theta=0.15$, with $\Delta Q_{\rm ideal}=\HwThetaAIdeal$\,\%: \qtrace{} flags \HwThetaAQT{} of \HwThetaAN{} link-epochs, the per-basis QBER test \HwThetaAQB{} and the pooled QBER test \HwThetaAQP. The pooled test loses power because the probe disturbs one basis at a time and pooling dilutes the signal. The per-basis test is close to optimal here, since on qubits almost every disturbance is a change of an error rate; \qtrace{} matches it while retaining channels for asymmetric and unsifted deviations.

\subsection{Rounds to detection and information leaked before the alarm}\label{sec:hw-info}

\begin{figure}[!tbp]
\centering
\includegraphics[width=\textwidth]{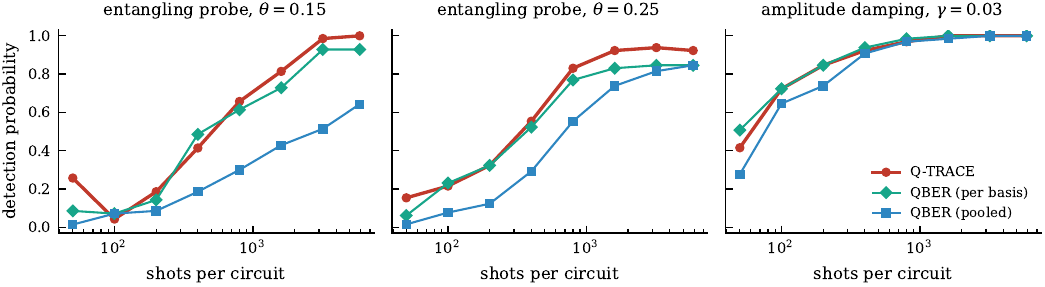}
\caption{Detection probability at $\alpha=10^{-3}$ against the number of shots per circuit of the tested epoch, for the three weakest probes on \HwBackend{} (condition-matched reference; mean over five random subsamples per point).}
\label{fig:rounds}
\end{figure}

Because every shot was recorded, detection can be studied as a function of the number of rounds by subsampling the shots of the tested epoch. \Cref{fig:rounds} shows the detection probability against shots per circuit for the three weakest probes, and \cref{tab:hwleak} the number of shots per circuit at which half of the attacked link-epochs are flagged. For $\theta=0.15$, \qtrace{} needs \HwShotsWeakQT{} shots, the per-basis test \HwShotsWeakQB{} and the pooled QBER test \HwShotsWeakQP, a factor of \HwShotsRatio{} more. The omnibus channel alone would need \HwShotsWeakOM.

Eve's ancilla measurements turn this into an operational quantity. On sifted rounds in which her probe basis matched Alice's basis, her measured information agrees with \cref{eq:infodisturb} up to the effect of her own readout errors (\cref{fig:hwphys}b and \cref{tab:hwinfo}); for $\theta=0.40$ it is \HwInfoFourMeas{} bit against \HwInfoFourIdeal{} ideally, and for the intercept--resend limit $\theta=\pi/2$ it is \HwIRInfo{} bit with a guessing error of \HwIRGuess. Multiplying the information per sifted bit by the number of sifted rounds needed for detection gives the information Eve gathers before the median first alarm: for $\theta=0.15$ this is \HwLeakWeakQT{} bits under \qtrace{}, \HwLeakWeakQB{} bits under the per-basis QBER test and \HwLeakWeakQP{} bits under the pooled QBER monitor. These are small numbers in absolute terms; their ratio is what transfers to a deployed system, where the same reasoning applies per window.

\section{Discussion}\label{sec:discussion}

\subsection{What the results establish}

Three findings carry over beyond the particular twin and processor used here. The first is that a physical model of the receiver is a better basis for a monitor than a learned model of attacks. A monitor that knows what an honest link can produce needs no attack data, controls its false-alarm rate through validation on honest links, treats loss as a nuisance rather than an anomaly, and reports which physical constraint an alarm violates. On the benchmark it detected more attacks than every comparison method in every attack group except the long-link regime, where the only method with a higher nominal rate achieved it by alarming on \FALongQBERstatic\,\% of honest long links.

The second finding is that detectability is a property of the attack and the telemetry, and can be computed. The exponent $\Dstar$ separates three situations that a test-set accuracy conflates: an attack that a monitor misses although the information is there, an attack that no monitor can catch within the available time, and an attack that leaves no trace at all. Reporting attack strength in units of $\Dstar$, or of the delay bound $\log(1/\alpha)/(N\Dstar)$, would make results from different systems comparable.

The third finding concerns practice. On real hardware, every monitor tested, including the plain QBER monitors that systems run today, raised alarms on honest links because the reference statistics had been acquired under different operating conditions. Monitoring is only as good as its reference, and the reference must be acquired in operation.

\subsection{Relation to QBER monitoring and to device-independent protocols}

On the qubit hardware, where nearly every disturbance shows up as a change of an error rate, the per-basis QBER test was close to optimal and \qtrace{} matched it. On the optical twin the situation is different: loss, multi-photon pulses, two detectors and three intensities give the receiver statistics a structure that QBER summarises poorly, and the attacks that matter most in practice (PNS, blinding, time-shift, randomness failures) exploit exactly that structure. The advantage of the honest-manifold approach therefore grows with the physical richness of the telemetry. Measurement-device-independent protocols \cite{LoCurtyQi2012} remove detector side channels by construction; for the large installed base of prepare-and-measure systems, a monitor of the kind described here is a low-cost complement that needs no hardware change.

\subsection{Deployment considerations}

The computational cost is modest: calibration takes well under a second per link and one monitoring step requires a one-dimensional Newton iteration and a few small matrix operations per window. The commissioning windows must be attack-free, which is the same trust assumption made when a system is first calibrated. Re-commissioning after maintenance is necessary, and the moving-reference channels make the monitor tolerant to slow drift between commissionings. An alarm need not abort the protocol; it can be used to exclude the affected block from key distillation, as proposed in the operational framework of Mohamed and Al-Kuwari \cite{MohamedAlKuwari2026}. Finite-size key rates \cite{Tomamichel2012,Lim2014} are computed from the same statistics and can be combined with the alarm decision.

\subsection{Limitations}\label{sec:limitations}

The study has limitations that bound its claims. (i) The simulation results depend on the twin and its parameter ranges, which we chose to be realistic but which cannot represent every system; the twin contains mismatch effects but is still a model. (ii) False-alarm control is empirical, through validation quantiles, rather than exact; windows within a link are not independent, and the run-level false-alarm rate depends on the monitoring horizon. (iii) The detectability bound assumes a stationary honest link and is asymptotic; under drift the achievable delay is larger. (iv) The attacks are physically motivated but not adaptive: an eavesdropper who knows the monitor would minimise $\Dstar$, and the space of such attacks has not been explored. (v) The hardware study emulates the channel with superconducting qubits, used a single processor and a single run, has 12 to 16 link-epochs per probe, and its condition-matched analysis was defined after the pre-registered one failed. A follow-up run with timing-matched commissioning circuits is needed to confirm the matched results prospectively. (vi) The afterpulse probability is taken from a specification; in systems where it is unknown it must be calibrated jointly with the background, which is poorly identified without loss variation during commissioning.

\section{Conclusion}\label{sec:conclusion}

We have shown that implementation attacks on decoy-state QKD can be detected without attack data by testing receiver telemetry against the manifold of statistics that an honest, lossy link can produce. The construction combines an exact count-level receiver model, an efficient treatment of the transmittance nuisance, physically structured score tests and drift-robust references. It comes with a detector-independent exponent that bounds what any telemetry monitor can achieve and that identifies attacks no monitor can see. On a benchmark with deliberate model mismatch the monitor detected substantially more attacks than industry monitors, unsupervised detectors and a supervised classifier at the same false-alarm rate, with delays close to the information-theoretic bound. A hardware study with physically executed eavesdroppers confirmed the probe physics, demonstrated low-rate attack detection with a condition-matched reference, and showed that a reference acquired under different operating conditions defeats every monitor, including the QBER monitors in use today. Natural next steps are an extension of the bound to drifting links, the study of adaptive attacks that minimise $\Dstar$, finite-key integration, and validation on an optical testbed.

\section*{Declarations}
\addcontentsline{toc}{section}{Declarations}

\paragraph{Funding.} No funding was received for this work.

\paragraph{Competing interests.} The author declares no competing interests.

\paragraph{Data availability.} The raw per-shot hardware data, all simulation results and the summary tables from which every number in this paper is generated (directories \path{results/hardware}, \path{results/benchmark} and \path{results/summary}) are available at \url{https://github.com/Sumitchongder/Q-TRACE}.

\paragraph{Code availability.} The complete source code, configuration files, the hardware notebook and the scripts that regenerate all figures and tables are available at \url{https://github.com/Sumitchongder/Q-TRACE} under the MIT licence.

\paragraph{Author contributions.} S.T.C. conceived the study, developed the methods and software, performed the simulations and the hardware experiment, analysed the data and wrote the manuscript.

\paragraph{Use of generative AI.} A large language model (Claude, Anthropic) was used to assist with software development, data-analysis code and the drafting and editing of the manuscript. The author reviewed, verified and takes full responsibility for all content, including the methods, results and their interpretation.

\paragraph{Acknowledgements.} We acknowledge the use of IBM Quantum services for this work. The views expressed are those of the author and do not reflect the official policy or position of IBM or the IBM Quantum team.

\bibliographystyle{unsrtnat}
\bibliography{references}

@inproceedings{BB84,
  author    = {Bennett, Charles H. and Brassard, Gilles},
  title     = {Quantum cryptography: Public key distribution and coin tossing},
  booktitle = {Proceedings of the IEEE International Conference on Computers, Systems and Signal Processing},
  address   = {Bangalore, India},
  pages     = {175--179},
  year      = {1984}
}

@article{Ekert1991,
  author  = {Ekert, Artur K.},
  title   = {Quantum cryptography based on {B}ell's theorem},
  journal = {Physical Review Letters}, volume = {67}, pages = {661--663}, year = {1991},
  doi     = {10.1103/PhysRevLett.67.661}
}

@article{ShorPreskill2000,
  author  = {Shor, Peter W. and Preskill, John},
  title   = {Simple proof of security of the {BB84} quantum key distribution protocol},
  journal = {Physical Review Letters}, volume = {85}, pages = {441--444}, year = {2000},
  doi     = {10.1103/PhysRevLett.85.441}
}

@article{GLLP2004,
  author  = {Gottesman, Daniel and Lo, Hoi-Kwong and L{\"u}tkenhaus, Norbert and Preskill, John},
  title   = {Security of quantum key distribution with imperfect devices},
  journal = {Quantum Information and Computation}, volume = {4}, number = {5}, pages = {325--360}, year = {2004}
}

@article{Scarani2009,
  author  = {Scarani, Valerio and Bechmann-Pasquinucci, Helle and Cerf, Nicolas J. and Du{\v{s}}ek, Miloslav and L{\"u}tkenhaus, Norbert and Peev, Momtchil},
  title   = {The security of practical quantum key distribution},
  journal = {Reviews of Modern Physics}, volume = {81}, pages = {1301--1350}, year = {2009},
  doi     = {10.1103/RevModPhys.81.1301}
}

@article{Xu2020,
  author  = {Xu, Feihu and Ma, Xiongfeng and Zhang, Qiang and Lo, Hoi-Kwong and Pan, Jian-Wei},
  title   = {Secure quantum key distribution with realistic devices},
  journal = {Reviews of Modern Physics}, volume = {92}, pages = {025002}, year = {2020},
  doi     = {10.1103/RevModPhys.92.025002}
}

@article{Pirandola2020,
  author  = {Pirandola, Stefano and Andersen, Ulrik L. and Banchi, Leonardo and Berta, Mario and Bunandar, Darius and Colbeck, Roger and Englund, Dirk and Gehring, Tobias and Lupo, Cosmo and Ottaviani, Carlo and Pereira, Jason L. and Razavi, Mohsen and Shamsul Shaari, Jesni and Tomamichel, Marco and Usenko, Vladyslav C. and Vallone, Giuseppe and Villoresi, Paolo and Wallden, Petros},
  title   = {Advances in quantum cryptography},
  journal = {Advances in Optics and Photonics}, volume = {12}, number = {4}, pages = {1012--1236}, year = {2020},
  doi     = {10.1364/AOP.361502}
}

@article{Diamanti2016,
  author  = {Diamanti, Eleni and Lo, Hoi-Kwong and Qi, Bing and Yuan, Zhiliang},
  title   = {Practical challenges in quantum key distribution},
  journal = {npj Quantum Information}, volume = {2}, pages = {16025}, year = {2016},
  doi     = {10.1038/npjqi.2016.25}
}

@article{Hwang2003,
  author  = {Hwang, Won-Young},
  title   = {Quantum key distribution with high loss: Toward global secure communication},
  journal = {Physical Review Letters}, volume = {91}, pages = {057901}, year = {2003},
  doi     = {10.1103/PhysRevLett.91.057901}
}

@article{LoMaChen2005,
  author  = {Lo, Hoi-Kwong and Ma, Xiongfeng and Chen, Kai},
  title   = {Decoy state quantum key distribution},
  journal = {Physical Review Letters}, volume = {94}, pages = {230504}, year = {2005},
  doi     = {10.1103/PhysRevLett.94.230504}
}

@article{Wang2005,
  author  = {Wang, Xiang-Bin},
  title   = {Beating the photon-number-splitting attack in practical quantum cryptography},
  journal = {Physical Review Letters}, volume = {94}, pages = {230503}, year = {2005},
  doi     = {10.1103/PhysRevLett.94.230503}
}

@article{Ma2005,
  author  = {Ma, Xiongfeng and Qi, Bing and Zhao, Yi and Lo, Hoi-Kwong},
  title   = {Practical decoy state for quantum key distribution},
  journal = {Physical Review A}, volume = {72}, pages = {012326}, year = {2005},
  doi     = {10.1103/PhysRevA.72.012326}
}

@article{LoChauArdehali2005,
  author  = {Lo, Hoi-Kwong and Chau, Hoi Fung and Ardehali, M.},
  title   = {Efficient quantum key distribution scheme and a proof of its unconditional security},
  journal = {Journal of Cryptology}, volume = {18}, pages = {133--165}, year = {2005}
}

@article{Tomamichel2012,
  author  = {Tomamichel, Marco and Lim, Charles Ci Wen and Gisin, Nicolas and Renner, Renato},
  title   = {Tight finite-key analysis for quantum cryptography},
  journal = {Nature Communications}, volume = {3}, pages = {634}, year = {2012},
  doi     = {10.1038/ncomms1631}
}

@article{Lim2014,
  author  = {Lim, Charles Ci Wen and Curty, Marcos and Walenta, Nino and Xu, Feihu and Zbinden, Hugo},
  title   = {Concise security bounds for practical decoy-state quantum key distribution},
  journal = {Physical Review A}, volume = {89}, pages = {022307}, year = {2014},
  doi     = {10.1103/PhysRevA.89.022307}
}

@article{LoCurtyQi2012,
  author  = {Lo, Hoi-Kwong and Curty, Marcos and Qi, Bing},
  title   = {Measurement-device-independent quantum key distribution},
  journal = {Physical Review Letters}, volume = {108}, pages = {130503}, year = {2012},
  doi     = {10.1103/PhysRevLett.108.130503}
}

@article{Acin2007,
  author  = {Ac{\'\i}n, Antonio and Brunner, Nicolas and Gisin, Nicolas and Massar, Serge and Pironio, Stefano and Scarani, Valerio},
  title   = {Device-independent security of quantum cryptography against collective attacks},
  journal = {Physical Review Letters}, volume = {98}, pages = {230501}, year = {2007},
  doi     = {10.1103/PhysRevLett.98.230501}
}

@article{Liao2017,
  author  = {Liao, Sheng-Kai and others},
  title   = {Satellite-to-ground quantum key distribution},
  journal = {Nature}, volume = {549}, pages = {43--47}, year = {2017},
  doi     = {10.1038/nature23655}
}

@article{Chen2021,
  author  = {Chen, Yu-Ao and others},
  title   = {An integrated space-to-ground quantum communication network over 4,600 kilometres},
  journal = {Nature}, volume = {589}, pages = {214--219}, year = {2021},
  doi     = {10.1038/s41586-020-03093-8}
}

@article{Brassard2000,
  author  = {Brassard, Gilles and L{\"u}tkenhaus, Norbert and Mor, Tal and Sanders, Barry C.},
  title   = {Limitations on practical quantum cryptography},
  journal = {Physical Review Letters}, volume = {85}, pages = {1330--1333}, year = {2000},
  doi     = {10.1103/PhysRevLett.85.1330}
}

@article{Lutkenhaus2000,
  author  = {L{\"u}tkenhaus, Norbert},
  title   = {Security against individual attacks for realistic quantum key distribution},
  journal = {Physical Review A}, volume = {61}, pages = {052304}, year = {2000},
  doi     = {10.1103/PhysRevA.61.052304}
}

@article{Lydersen2010,
  author  = {Lydersen, Lars and Wiechers, Carlos and Wittmann, Christoffer and Elser, Dominique and Skaar, Johannes and Makarov, Vadim},
  title   = {Hacking commercial quantum cryptography systems by tailored bright illumination},
  journal = {Nature Photonics}, volume = {4}, pages = {686--689}, year = {2010},
  doi     = {10.1038/nphoton.2010.214}
}

@article{Lydersen2011superlinear,
  author  = {Lydersen, Lars and Jain, Nitin and Wittmann, Christoffer and Mar{\o}y, {\O}ystein and Skaar, Johannes and Marquardt, Christoph and Makarov, Vadim and Leuchs, Gerd},
  title   = {Superlinear threshold detectors in quantum cryptography},
  journal = {Physical Review A}, volume = {84}, pages = {032320}, year = {2011},
  doi     = {10.1103/PhysRevA.84.032320}
}

@article{Gerhardt2011,
  author  = {Gerhardt, Ilja and Liu, Qin and Lamas-Linares, Ant{\'\i}a and Skaar, Johannes and Kurtsiefer, Christian and Makarov, Vadim},
  title   = {Full-field implementation of a perfect eavesdropper on a quantum cryptography system},
  journal = {Nature Communications}, volume = {2}, pages = {349}, year = {2011},
  doi     = {10.1038/ncomms1348}
}

@article{Wiechers2011,
  author  = {Wiechers, Carlos and Lydersen, Lars and Wittmann, Christoffer and Elser, Dominique and Skaar, Johannes and Marquardt, Christoph and Makarov, Vadim and Leuchs, Gerd},
  title   = {After-gate attack on a quantum cryptosystem},
  journal = {New Journal of Physics}, volume = {13}, pages = {013043}, year = {2011},
  doi     = {10.1088/1367-2630/13/1/013043}
}

@article{Makarov2006,
  author  = {Makarov, Vadim and Anisimov, Andrey and Skaar, Johannes},
  title   = {Effects of detector efficiency mismatch on security of quantum cryptosystems},
  journal = {Physical Review A}, volume = {74}, pages = {022313}, year = {2006},
  doi     = {10.1103/PhysRevA.74.022313}
}

@article{Qi2007,
  author  = {Qi, Bing and Fung, Chi-Hang Fred and Lo, Hoi-Kwong and Ma, Xiongfeng},
  title   = {Time-shift attack in practical quantum cryptosystems},
  journal = {Quantum Information and Computation}, volume = {7}, number = {1}, pages = {73--82}, year = {2007}
}

@article{Zhao2008,
  author  = {Zhao, Yi and Fung, Chi-Hang Fred and Qi, Bing and Chen, Christine and Lo, Hoi-Kwong},
  title   = {Quantum hacking: Experimental demonstration of time-shift attack against practical quantum-key-distribution systems},
  journal = {Physical Review A}, volume = {78}, pages = {042333}, year = {2008},
  doi     = {10.1103/PhysRevA.78.042333}
}

@article{Gisin2006THA,
  author  = {Gisin, Nicolas and Fasel, Sylvain and Kraus, Barbara and Zbinden, Hugo and Ribordy, Gr{\'e}goire},
  title   = {Trojan-horse attacks on quantum-key-distribution systems},
  journal = {Physical Review A}, volume = {73}, pages = {022320}, year = {2006},
  doi     = {10.1103/PhysRevA.73.022320}
}

@article{Lucamarini2015,
  author  = {Lucamarini, Marco and Choi, Iris and Ward, Martin B. and Dynes, James F. and Yuan, Zhiliang L. and Shields, Andrew J.},
  title   = {Practical security bounds against the {T}rojan-horse attack in quantum key distribution},
  journal = {Physical Review X}, volume = {5}, pages = {031030}, year = {2015},
  doi     = {10.1103/PhysRevX.5.031030}
}

@article{Jain2014THA,
  author  = {Jain, Nitin and Anisimova, Elena and Khan, Imran and Makarov, Vadim and Marquardt, Christoph and Leuchs, Gerd},
  title   = {Trojan-horse attacks threaten the security of practical quantum cryptography},
  journal = {New Journal of Physics}, volume = {16}, pages = {123030}, year = {2014},
  doi     = {10.1088/1367-2630/16/12/123030}
}

@article{Jain2011calibration,
  author  = {Jain, Nitin and Wittmann, Christoffer and Lydersen, Lars and Wiechers, Carlos and Elser, Dominique and Marquardt, Christoph and Makarov, Vadim and Leuchs, Gerd},
  title   = {Device calibration impacts security of quantum key distribution},
  journal = {Physical Review Letters}, volume = {107}, pages = {110501}, year = {2011},
  doi     = {10.1103/PhysRevLett.107.110501}
}

@article{Jain2016review,
  author  = {Jain, Nitin and Stiller, Birgit and Khan, Imran and Elser, Dominique and Marquardt, Christoph and Leuchs, Gerd},
  title   = {Attacks on practical quantum key distribution systems (and how to prevent them)},
  journal = {Contemporary Physics}, volume = {57}, number = {3}, pages = {366--387}, year = {2016}
}

@article{XuQiLo2010,
  author  = {Xu, Feihu and Qi, Bing and Lo, Hoi-Kwong},
  title   = {Experimental demonstration of phase-remapping attack in a practical quantum key distribution system},
  journal = {New Journal of Physics}, volume = {12}, pages = {113026}, year = {2010},
  doi     = {10.1088/1367-2630/12/11/113026}
}

@article{Bugge2014,
  author  = {Bugge, Audun Nystad and Sauge, Sebastien and Ghazali, Aina Mardhiyah M. and Skaar, Johannes and Lydersen, Lars and Makarov, Vadim},
  title   = {Laser damage helps the eavesdropper in quantum cryptography},
  journal = {Physical Review Letters}, volume = {112}, pages = {070503}, year = {2014},
  doi     = {10.1103/PhysRevLett.112.070503}
}

@article{Makarov2016,
  author  = {Makarov, Vadim and Bourgoin, Jean-Philippe and Chaiwongkhot, Poompong and Gagn{\'e}, Mathieu and Jennewein, Thomas and Kaiser, Sarah and Kashyap, Raman and Legr{\'e}, Matthieu and Minshull, Carter and Sajeed, Shihan},
  title   = {Creation of backdoors in quantum communications via laser damage},
  journal = {Physical Review A}, volume = {94}, pages = {030302}, year = {2016},
  doi     = {10.1103/PhysRevA.94.030302}
}

@article{Huang2019seeding,
  author  = {Huang, Anqi and Navarrete, {\'A}lvaro and Sun, Shi-Hai and Chaiwongkhot, Poompong and Curty, Marcos and Makarov, Vadim},
  title   = {Laser-seeding attack in quantum key distribution},
  journal = {Physical Review Applied}, volume = {12}, pages = {064043}, year = {2019},
  doi     = {10.1103/PhysRevApplied.12.064043}
}

@article{Ponosova2022,
  author  = {Ponosova, Anastasiya and Ruzhitskaya, Daria and Chaiwongkhot, Poompong and Egorov, Vladimir and Makarov, Vadim and Huang, Anqi},
  title   = {Protecting fiber-optic quantum key distribution sources against light-injection attacks},
  journal = {PRX Quantum}, volume = {3}, pages = {040307}, year = {2022},
  doi     = {10.1103/PRXQuantum.3.040307}
}

@article{SunHuang2022,
  author  = {Sun, Shihai and Huang, Anqi},
  title   = {A review of security evaluation of practical quantum key distribution system},
  journal = {Entropy}, volume = {24}, pages = {260}, year = {2022},
  doi     = {10.3390/e24020260}
}

@article{Pereira2020,
  author  = {Pereira, Margarida and Kato, Go and Mizutani, Akihiro and Curty, Marcos and Tamaki, Kiyoshi},
  title   = {Quantum key distribution with correlated sources},
  journal = {Science Advances}, volume = {6}, pages = {eaaz4487}, year = {2020},
  doi     = {10.1126/sciadv.aaz4487}
}

@article{Tamaki2016,
  author  = {Tamaki, Kiyoshi and Curty, Marcos and Lucamarini, Marco},
  title   = {Decoy-state quantum key distribution with a leaky source},
  journal = {New Journal of Physics}, volume = {18}, pages = {065008}, year = {2016},
  doi     = {10.1088/1367-2630/18/6/065008}
}

@article{Zapatero2021,
  author  = {Zapatero, V{\'\i}ctor and Navarrete, {\'A}lvaro and Tamaki, Kiyoshi and Curty, Marcos},
  title   = {Security of quantum key distribution with intensity correlations},
  journal = {Quantum}, volume = {5}, pages = {602}, year = {2021},
  doi     = {10.22331/q-2021-12-07-602}
}

@article{Fuchs1997,
  author  = {Fuchs, Christopher A. and Gisin, Nicolas and Griffiths, Robert B. and Niu, Chi-Sheng and Peres, Asher},
  title   = {Optimal eavesdropping in quantum cryptography. {I}. {I}nformation bound and optimal strategy},
  journal = {Physical Review A}, volume = {56}, pages = {1163--1172}, year = {1997},
  doi     = {10.1103/PhysRevA.56.1163}
}

@book{Helstrom1976,
  author    = {Helstrom, Carl W.},
  title     = {Quantum Detection and Estimation Theory},
  publisher = {Academic Press}, address = {New York}, year = {1976}
}

@article{Page1954,
  author  = {Page, E. S.},
  title   = {Continuous inspection schemes},
  journal = {Biometrika}, volume = {41}, pages = {100--115}, year = {1954},
  doi     = {10.1093/biomet/41.1-2.100}
}

@article{Lorden1971,
  author  = {Lorden, Gary},
  title   = {Procedures for reacting to a change in distribution},
  journal = {The Annals of Mathematical Statistics}, volume = {42}, number = {6}, pages = {1897--1908}, year = {1971},
  doi     = {10.1214/aoms/1177693055}
}

@article{Moustakides1986,
  author  = {Moustakides, George V.},
  title   = {Optimal stopping times for detecting changes in distributions},
  journal = {The Annals of Statistics}, volume = {14}, number = {4}, pages = {1379--1387}, year = {1986},
  doi     = {10.1214/aos/1176350164}
}

@article{Lai1998,
  author  = {Lai, Tze Leung},
  title   = {Information bounds and quick detection of parameter changes in stochastic systems},
  journal = {IEEE Transactions on Information Theory}, volume = {44}, number = {7}, pages = {2917--2929}, year = {1998},
  doi     = {10.1109/18.737522}
}

@book{Basseville1993,
  author    = {Basseville, Mich{\`e}le and Nikiforov, Igor V.},
  title     = {Detection of Abrupt Changes: Theory and Application},
  publisher = {Prentice Hall}, address = {Englewood Cliffs, NJ}, year = {1993}
}

@book{Tartakovsky2014,
  author    = {Tartakovsky, Alexander and Nikiforov, Igor and Basseville, Mich{\`e}le},
  title     = {Sequential Analysis: Hypothesis Testing and Changepoint Detection},
  publisher = {CRC Press}, address = {Boca Raton, FL}, year = {2014}
}

@article{Rao1948,
  author  = {Rao, C. Radhakrishna},
  title   = {Large sample tests of statistical hypotheses concerning several parameters with applications to problems of estimation},
  journal = {Mathematical Proceedings of the Cambridge Philosophical Society}, volume = {44}, number = {1}, pages = {50--57}, year = {1948},
  doi     = {10.1017/S0305004100023987}
}

@incollection{Neyman1959,
  author    = {Neyman, Jerzy},
  title     = {Optimal asymptotic tests of composite statistical hypotheses},
  booktitle = {Probability and Statistics: The Harald Cram{\'e}r Volume},
  editor    = {Grenander, Ulf},
  publisher = {Wiley}, address = {New York}, pages = {213--234}, year = {1959}
}

@article{FreemanTukey1950,
  author  = {Freeman, Murray F. and Tukey, John W.},
  title   = {Transformations related to the angular and the square root},
  journal = {The Annals of Mathematical Statistics}, volume = {21}, number = {4}, pages = {607--611}, year = {1950},
  doi     = {10.1214/aoms/1177729756}
}

@article{KullbackLeibler1951,
  author  = {Kullback, S. and Leibler, R. A.},
  title   = {On information and sufficiency},
  journal = {The Annals of Mathematical Statistics}, volume = {22}, number = {1}, pages = {79--86}, year = {1951},
  doi     = {10.1214/aoms/1177729694}
}

@book{CoverThomas2006,
  author    = {Cover, Thomas M. and Thomas, Joy A.},
  title     = {Elements of Information Theory},
  edition   = {2nd}, publisher = {Wiley-Interscience}, address = {Hoboken, NJ}, year = {2006}
}

@book{LehmannRomano2005,
  author    = {Lehmann, Erich L. and Romano, Joseph P.},
  title     = {Testing Statistical Hypotheses},
  edition   = {3rd}, publisher = {Springer}, address = {New York}, year = {2005}
}

@article{LedoitWolf2004,
  author  = {Ledoit, Olivier and Wolf, Michael},
  title   = {A well-conditioned estimator for large-dimensional covariance matrices},
  journal = {Journal of Multivariate Analysis}, volume = {88}, number = {2}, pages = {365--411}, year = {2004},
  doi     = {10.1016/S0047-259X(03)00096-4}
}

@incollection{Miller1955,
  author    = {Miller, George A.},
  title     = {Note on the bias of information estimates},
  booktitle = {Information Theory in Psychology: Problems and Methods},
  editor    = {Quastler, Henry},
  publisher = {Free Press}, address = {Glencoe, IL}, pages = {95--100}, year = {1955}
}

@article{Mao2020,
  author  = {Mao, Yiyu and Huang, Weiti and Zhong, Hai and Wang, Yijun and Qin, Hao and Guo, Ying and Huang, Duan},
  title   = {Detecting quantum attacks: a machine learning based defense strategy for practical continuous-variable quantum key distribution},
  journal = {New Journal of Physics}, volume = {22}, pages = {083073}, year = {2020}
}

@article{Luo2022,
  author  = {Luo, Hao and Zhang, Ling and Qin, Hao and Sun, Shihai and Huang, Peng and Wang, Yijun and Wu, Zhengquan and Guo, Ying and Huang, Duan},
  title   = {Beyond universal attack detection for continuous-variable quantum key distribution via deep learning},
  journal = {Physical Review A}, volume = {105}, pages = {042411}, year = {2022},
  doi     = {10.1103/PhysRevA.105.042411}
}

@article{XuJ2024,
  author  = {Xu, Junxuan and Ma, Xiao and Liu, Jingyi and Zhang, Chunhui and Li, Hongwei and Zhou, Xingyu and Wang, Qin},
  title   = {Automatically identifying imperfections and attacks in practical quantum key distribution systems via machine learning},
  journal = {Science China Information Sciences}, volume = {67}, pages = {202501}, year = {2024}
}

@article{AlMohammed2021,
  author  = {Al-Mohammed, Hasan Abbas and Al-Ali, Afnan and Yaacoub, Elias and Qidwai, Uvais and Abualsaud, Khalid and Rzewuski, Stanis{\l}aw and Flizikowski, Adam},
  title   = {Machine learning techniques for detecting attackers during quantum key distribution in {IoT} networks with application to railway scenarios},
  journal = {IEEE Access}, volume = {9}, pages = {136994--137004}, year = {2021}
}

@article{Ren2021,
  author  = {Ren, Zhi-Ang and Chen, Yu-Peng and Liu, Jing-Yang and Ding, Hua-Jian and Wang, Qin},
  title   = {Implementation of machine learning in quantum key distributions},
  journal = {IEEE Communications Letters}, volume = {25}, number = {3}, pages = {940--944}, year = {2021}
}

@article{AlKuwari2025,
  author  = {Al-Kuwari, Ali and Mohamed, Noureldin and Al-Kuwari, Saif and Farouk, Ahmed and Behera, Bikash K.},
  title   = {Resisting quantum key distribution attacks using quantum machine learning},
  journal = {arXiv preprint}, year = {2025}, eprint = {2509.14282}, note = {arXiv:2509.14282}
}

@article{MohamedAlKuwari2026,
  author  = {Mohamed, Noureldin and Al-Kuwari, Saif},
  title   = {Adversarial learning game for intrusion detection in quantum key distribution},
  journal = {arXiv preprint}, year = {2026}, eprint = {2603.03502}, note = {arXiv:2603.03502}
}

@article{Liu2025AD,
  author  = {Liu, Junxuan and Huang, Bingcheng and Su, Jialei and Peng, Qingquan and Huang, Anqi},
  title   = {Deep anomaly detection for active attacks on the receiver in quantum key distribution},
  journal = {arXiv preprint}, year = {2025}, eprint = {2508.12749}, note = {arXiv:2508.12749}
}

@inproceedings{Liu2008IF,
  author    = {Liu, Fei Tony and Ting, Kai Ming and Zhou, Zhi-Hua},
  title     = {Isolation forest},
  booktitle = {Proceedings of the 8th IEEE International Conference on Data Mining},
  pages     = {413--422}, year = {2008}, doi = {10.1109/ICDM.2008.17}
}

@article{Scholkopf2001,
  author  = {Sch{\"o}lkopf, Bernhard and Platt, John C. and Shawe-Taylor, John and Smola, Alex J. and Williamson, Robert C.},
  title   = {Estimating the support of a high-dimensional distribution},
  journal = {Neural Computation}, volume = {13}, number = {7}, pages = {1443--1471}, year = {2001},
  doi     = {10.1162/089976601750264965}
}

@inproceedings{ChenGuestrin2016,
  author    = {Chen, Tianqi and Guestrin, Carlos},
  title     = {{XGBoost}: A scalable tree boosting system},
  booktitle = {Proceedings of the 22nd ACM SIGKDD International Conference on Knowledge Discovery and Data Mining},
  pages     = {785--794}, year = {2016}, doi = {10.1145/2939672.2939785}
}

@article{Pedregosa2011,
  author  = {Pedregosa, Fabian and others},
  title   = {Scikit-learn: Machine learning in {P}ython},
  journal = {Journal of Machine Learning Research}, volume = {12}, pages = {2825--2830}, year = {2011}
}

@article{Havlicek2019,
  author  = {Havl{\'\i}{\v{c}}ek, Vojt{\v{e}}ch and C{\'o}rcoles, Antonio D. and Temme, Kristan and Harrow, Aram W. and Kandala, Abhinav and Chow, Jerry M. and Gambetta, Jay M.},
  title   = {Supervised learning with quantum-enhanced feature spaces},
  journal = {Nature}, volume = {567}, pages = {209--212}, year = {2019},
  doi     = {10.1038/s41586-019-0980-2}
}

@article{SchuldKilloran2019,
  author  = {Schuld, Maria and Killoran, Nathan},
  title   = {Quantum machine learning in feature {H}ilbert spaces},
  journal = {Physical Review Letters}, volume = {122}, pages = {040504}, year = {2019},
  doi     = {10.1103/PhysRevLett.122.040504}
}

@article{Qiskit2024,
  author  = {Javadi-Abhari, Ali and Treinish, Matthew and Krsulich, Kevin and Wood, Christopher J. and Lishman, Jake and Gacon, Julien and Martiel, Simon and Nation, Paul D. and Bishop, Lev S. and Cross, Andrew W. and Johnson, Blake R. and Gambetta, Jay M.},
  title   = {Quantum computing with {Q}iskit},
  journal = {arXiv preprint}, year = {2024}, eprint = {2405.08810}, note = {arXiv:2405.08810}
}

\clearpage
\appendix
\section{Derivations}\label{app:derivations}

\subsection{Freeman--Tukey residuals}\label{app:ft}

Let $n\sim\mathrm{Poisson}(E)$. Freeman and Tukey \cite{FreemanTukey1950} showed that $\sqrt{n}+\sqrt{n+1}$ has variance $1+O(E^{-1})$ and mean $\sqrt{4E+1}+O(E^{-3/2})$, so the residual $r=\sqrt{n}+\sqrt{n+1}-\sqrt{4E+1}$ of \cref{eq:ft} is approximately standard normal for moderate $E$ and remains well behaved, with mean and variance close to their limits, down to $E$ of order one. For a cell with $E\to0$ the residual tends to $r=\sqrt{n}+\sqrt{n+1}-1$, which is zero with probability $1-O(E)$; such cells contribute little to either statistic unless the attack populates them, which is exactly the situation for, for example, double clicks generated by back-scattered light. Multinomial cells with small cell probability behave like Poisson cells; for the no-click cells with probability close to one we do not form residuals, because they are determined by the click cells of the same setting.

\subsection{Variance inflation from estimated bias}\label{app:inflation}

Let $r_\tau=\mu+\varepsilon_\tau$ with independent $\varepsilon_\tau$ of unit variance per component, and let $b=W^{-1}\sum_{w=1}^W r_w$ be estimated on $W$ commissioning windows independent of the $M$ monitoring windows. Then
\begin{equation}
\mathrm{Var}\Bigl(\sum_{\tau=1}^M (r_\tau-b)\Bigr)=M+\frac{M^2}{W}=M\Bigl(1+\frac{M}{W}\Bigr)
\end{equation}
per component, which is the normalisation in \cref{eq:shift,eq:scorestat}. The moving-reference difference of \cref{eq:moving} has, by the same argument with the reference block of length $R$ in place of the commissioning windows, variance $M(1+M/R)$.

\subsection{Fisher information of the multinomial outcome model}\label{app:fisher}

For one setting with $N_s$ pulses and outcome probabilities $p_o(\vartheta)$, the score is $\sum_o n_o\nabla\log p_o$. Since $\sum_o p_o=1$ for all $\vartheta$, $\sum_o p_o\nabla\log p_o=\nabla\sum_o p_o=0$, so the score equals $\sum_o (n_o-N_sp_o)\nabla\log p_o$, which is the form \eqref{eq:score}. Its covariance under the model is
\begin{equation}
\mathrm{Cov}\Bigl(\sum_o n_o g_o\Bigr)=N_s\Bigl[\sum_o p_og_og_o\tr-\Bigl(\sum_o p_og_o\Bigr)\Bigl(\sum_o p_og_o\Bigr)\tr\Bigr]=N_s\sum_o p_og_og_o\tr,
\end{equation}
with $g_o=\nabla\log p_o$, because the second term vanishes by the same identity. Summing over settings gives the information $I_t$ of \cref{sec:score}. The derivatives are evaluated by central finite differences with step $10^{-4}$ in the transformed coordinates of \cref{eq:directions}.

\subsection{Efficient score and the nuisance parameter}\label{app:efficient}

Partition the score as $(u_\psi,u_\eta)$ and the information accordingly. The efficient score $u_\psi-I_{\psi\eta}I_{\eta\eta}^{-1}u_\eta$ is the residual of the $L^2$ projection of $u_\psi$ onto $u_\eta$ and is therefore uncorrelated with $u_\eta$. A first-order error $\delta$ in the nuisance changes $u_\psi$ by $-I_{\psi\eta}\delta$ and $u_\eta$ by $-I_{\eta\eta}\delta$ to first order, so the efficient score is unchanged to first order; this is why profiling $\eta$ per window does not bias the statistic. The covariance of the efficient score is the Schur complement $I_{\psi\psi}-I_{\psi\eta}I_{\eta\eta}^{-1}I_{\eta\psi}$ \cite{Neyman1959,LehmannRomano2005}.

\FloatBarrier
\section{Twin and attack parameters}\label{app:params}

Table~\ref{tab:twin} lists the sampling ranges of every link parameter and nuisance process of the twin, and Table~\ref{tab:attackranges} the strength ranges of the six attack families.

\begin{table}[!htbp]
\centering
\caption{Sampling ranges of the twin. Ranges are uniform unless marked log-uniform. Effects marked $\dagger$ are not represented in the honest model of the monitor.}
\label{tab:twin}
\small
\begin{tabular}{@{}>{\raggedright\arraybackslash}p{5.0cm}>{\raggedright\arraybackslash}p{5.6cm}>{\raggedright\arraybackslash}p{4.4cm}@{}}
\toprule
Quantity & Range & Note \\
\midrule
Signal mean photon number $\mu_\mathrm{s}$ & 0.40--0.60 & \\
Decoy ratio $\mu_\mathrm{d}/\mu_\mathrm{s}$ & 0.15--0.30 & vacuum $\mu_\mathrm{v}=0$ \\
Intensity probabilities $(p_\mathrm{s},p_\mathrm{d},p_\mathrm{v})$ & (0.70, 0.20, 0.10) & bases and bits uniform \\
Fibre length & 5--60\,km (long links: 60--120\,km) & attenuation 0.18--0.22\,dB/km \\
Receiver efficiency $\eta_B$ & 0.10--0.50 (long links: 0.05--0.20) & \\
Optical misalignment $e_Z,e_X$ & 0.5--2\,\%, basis factor 0.8--1.25 & \\
Dark-count probability $p_{dj}$ & $2\times10^{-7}$--$5\times10^{-6}$ per gate & log-uniform, per detector \\
Relative efficiency $\kappa$ & 0.85--1.15 & \\
Afterpulse probability $p_{\rm ap}$ & 0--3\,\% & datasheet value known to $\pm10$\,\% \\
Intensity jitter$^\dagger$ & 0.2--1\,\% per window & independent per intensity \\
Transmittance random walk & innovation 0.1--0.6\,\% per window & AR(1), coefficient 0.995 \\
Periodic transmittance modulation & amplitude 0--4\,\%, period 150--600 windows & \\
Polarisation drift$^\dagger$ & innovation 0--$2\times10^{-4}$ per window & AR(1), coefficient 0.98 \\
Dark-count temperature modulation$^\dagger$ & amplitude 0--10\,\% & same period as above \\
Pulses per window $N$ & $10^7$ & 1\,s at 10\,MHz \\
Windows per link & 200, of which 60 commissioning & \\
\bottomrule
\end{tabular}
\end{table}

\begin{table}[!htbp]
\centering
\caption{Attack strength ranges (log-uniform unless stated). Onsets are uniform between 20 and 100 windows after commissioning.}
\label{tab:attackranges}
\small
\begin{tabular}{@{}lll@{}}
\toprule
Attack & Strength parameter & Range \\
\midrule
Intercept--resend & intercepted fraction $f$ & 0.002--0.2 \\
Photon-number splitting & attacked fraction $f$ & 0.01--1 \\
Detector blinding & blinded fraction $f$ & 0.01--1 \\
Time-shift & shifted fraction $f$; mismatch $r$ & 0.02--1; $r$ uniform 0.1--0.5 \\
Trojan horse & back-scatter probability $q_{\rm bs}$ per gate & $10^{-8}$--$3\times10^{-6}$ \\
RNG bias & amplitude $\delta$; period $T$; phase $\phi$ & $5\times10^{-5}$--$2\times10^{-3}$; $T$ uniform 10--80; $\phi$ uniform \\
\bottomrule
\end{tabular}
\end{table}

\FloatBarrier
\section{Supplementary simulation results}\label{app:results}

\begin{table}[!htbp]
\centering
\caption{Window-level AUROC by attack group (attack-active windows against windows of attack-free test links). Mean\,$\pm$\,95\,\% confidence interval over held-out seeds.}
\label{tab:auroc}
\small
\begin{tabular}{lcccc}
\toprule
Method & Known & Zero-day & Composite & Long links \\
\midrule
\textbf{Q-TRACE} & \textbf{0.859\,$\pm$\,0.033} & \textbf{0.636\,$\pm$\,0.032} & \textbf{0.858\,$\pm$\,0.034} & \textbf{0.699\,$\pm$\,0.032} \\
XGBoost (sup.) & 0.848\,$\pm$\,0.026 & 0.499\,$\pm$\,0.043 & 0.771\,$\pm$\,0.041 & 0.718\,$\pm$\,0.041 \\
QBER CUSUM & 0.664\,$\pm$\,0.043 & 0.533\,$\pm$\,0.045 & 0.664\,$\pm$\,0.034 & 0.624\,$\pm$\,0.019 \\
Decoy CUSUM & 0.746\,$\pm$\,0.022 & 0.546\,$\pm$\,0.031 & 0.724\,$\pm$\,0.067 & 0.660\,$\pm$\,0.036 \\
QBER static & 0.567\,$\pm$\,0.064 & 0.487\,$\pm$\,0.072 & 0.545\,$\pm$\,0.065 & 0.571\,$\pm$\,0.075 \\
Autoencoder & 0.800\,$\pm$\,0.036 & 0.573\,$\pm$\,0.017 & 0.766\,$\pm$\,0.020 & 0.663\,$\pm$\,0.036 \\
Isolation forest & 0.767\,$\pm$\,0.033 & 0.554\,$\pm$\,0.014 & 0.745\,$\pm$\,0.049 & 0.677\,$\pm$\,0.033 \\
One-class SVM & 0.814\,$\pm$\,0.021 & 0.571\,$\pm$\,0.027 & 0.775\,$\pm$\,0.029 & 0.677\,$\pm$\,0.026 \\
QK one-class SVM & 0.523\,$\pm$\,0.018 & 0.498\,$\pm$\,0.013 & 0.521\,$\pm$\,0.014 & 0.501\,$\pm$\,0.025 \\
\bottomrule
\end{tabular}

\end{table}

\begin{table}[!htbp]
\centering
\caption{As \cref{tab:detection}, at a per-window false-alarm rate of $10^{-2}$.}
\label{tab:detection2}
\small
\resizebox{\textwidth}{!}{\begin{tabular}{lccccccc}
\toprule
Method & Known & Zero-day & Composite & Long links & FA honest & FA loss & FA long \\
\midrule
\textbf{Q-TRACE} & \textbf{73.7\,$\pm$\,7.8} & \textbf{42.3\,$\pm$\,4.0} & \textbf{74.7\,$\pm$\,5.9} & \textbf{49.3\,$\pm$\,9.6} & \textbf{33.2\,$\pm$\,20.5} & \textbf{16.8\,$\pm$\,11.3} & \textbf{20.8\,$\pm$\,18.4} \\
XGBoost (sup.) & 61.7\,$\pm$\,12.5 & 3.0\,$\pm$\,3.7 & 50.0\,$\pm$\,8.4 & 31.6\,$\pm$\,12.4 & 4.4\,$\pm$\,3.2 & 29.6\,$\pm$\,11.4 & 0.8\,$\pm$\,2.2 \\
QBER CUSUM & 31.2\,$\pm$\,4.0 & 4.0\,$\pm$\,3.3 & 23.7\,$\pm$\,6.3 & 12.9\,$\pm$\,4.9 & 4.8\,$\pm$\,5.7 & 4.8\,$\pm$\,5.4 & 0.0\,$\pm$\,0.0 \\
Decoy CUSUM & 5.0\,$\pm$\,1.0 & 0.0\,$\pm$\,0.0 & 2.0\,$\pm$\,2.3 & 0.9\,$\pm$\,1.5 & 0.0\,$\pm$\,0.0 & 17.6\,$\pm$\,5.7 & 0.0\,$\pm$\,0.0 \\
QBER static & 15.4\,$\pm$\,4.8 & 3.6\,$\pm$\,3.3 & 15.7\,$\pm$\,9.3 & 56.0\,$\pm$\,3.6 & 5.6\,$\pm$\,2.1 & 3.2\,$\pm$\,6.5 & 48.0\,$\pm$\,14.5 \\
Autoencoder & 6.3\,$\pm$\,3.6 & 0.0\,$\pm$\,0.0 & 4.0\,$\pm$\,3.1 & 1.8\,$\pm$\,3.0 & 0.0\,$\pm$\,0.0 & 20.8\,$\pm$\,16.2 & 0.0\,$\pm$\,0.0 \\
Isolation forest & 7.0\,$\pm$\,6.6 & 1.1\,$\pm$\,1.9 & 7.3\,$\pm$\,10.4 & 0.9\,$\pm$\,1.5 & 0.8\,$\pm$\,2.2 & 40.0\,$\pm$\,12.7 & 0.0\,$\pm$\,0.0 \\
One-class SVM & 0.0\,$\pm$\,0.0 & 0.0\,$\pm$\,0.0 & 0.0\,$\pm$\,0.0 & 0.0\,$\pm$\,0.0 & 0.0\,$\pm$\,0.0 & 0.0\,$\pm$\,0.0 & 0.0\,$\pm$\,0.0 \\
QK one-class SVM & 47.6\,$\pm$\,11.4 & 50.9\,$\pm$\,4.1 & 44.7\,$\pm$\,9.8 & 50.7\,$\pm$\,11.3 & 73.2\,$\pm$\,5.7 & 72.8\,$\pm$\,12.4 & 71.2\,$\pm$\,5.4 \\
\bottomrule
\end{tabular}
}
\end{table}

\begin{table}[!htbp]
\centering
\caption{Test-split attacks at a per-window false-alarm rate of $10^{-3}$: percentage of runs detected, median delay of detected runs, and mean key generated between onset and first alarm (or end of run).}
\label{tab:delaykey}
\small
\begin{tabular}{lccc}
\toprule
Method & Runs detected (\%) & Median delay (windows) & Key exposed (Mbit) \\
\midrule
\textbf{Q-TRACE} & \textbf{50.1} & \textbf{0} & \textbf{6.10} \\
XGBoost (sup.) & 29.6 & 5 & 9.70 \\
QBER CUSUM & 15.8 & 14 & 11.07 \\
Decoy CUSUM & 1.0 & 74 & 12.13 \\
QBER static & 8.4 & 0 & 11.70 \\
Autoencoder & 1.5 & 0 & 12.10 \\
Isolation forest & 2.7 & 38 & 11.88 \\
One-class SVM & 0.0 & -- & 12.14 \\
QK one-class SVM & 5.8 & 44 & 11.81 \\
\bottomrule
\end{tabular}

\end{table}

The key exposed is the key that the decoy-state analysis of \cref{eq:rate} would certify between the onset of an attack and the first alarm. It weights early detection of attacks on high-rate links, and it is the quantity that the security of a deployed system depends on when alarms are used to discard blocks.

\FloatBarrier
\section{Hardware details}\label{app:hardware}

\begin{table}[!htbp]
\centering
\caption{Calibration of the \HwLinks{} links on \HwBackend{} at the time of the run (backend-reported values).}
\label{tab:hwcal}
\small
\begin{tabular}{lccc}
\toprule
Quantity & Median & Min & Max \\
\midrule
CZ error (link pair) (\%) & 0.26 & 0.12 & 0.60 \\
Readout error, data qubit (\%) & 0.74 & 0.29 & 3.31 \\
Readout error, ancilla (\%) & 2.00 & 0.59 & 4.77 \\
$T_1$, data qubit ($\mu$s) & 164 & 68 & 364 \\
$T_2$, data qubit ($\mu$s) & 76 & 18 & 375 \\
$T_1$, ancilla ($\mu$s) & 170 & 98 & 440 \\
$T_2$, ancilla ($\mu$s) & 113 & 26 & 366 \\
\bottomrule
\end{tabular}

\end{table}

The commissioning QBER of the links ranged from \HwQberCalMin\,\% to \HwQberCalMax\,\% with a median of \HwQberCalMed\,\%. The highest value belongs to a link whose readout had degraded since the backend's last calibration; it was retained, because a monitor calibrated per link must cope with such links.

\begin{table}[!htbp]
\centering
\caption{Pre-registered analysis (reference: commissioning passes). Fraction of link-epochs flagged at $\alpha=10^{-3}$. Every test flags most honest link-epochs; see \cref{sec:hw-pre}.}
\label{tab:hwpre}
\small
\begin{tabular}{lccccc}
\toprule
Probe & $n$ & $\Delta Q_{\rm meas}$ (\%) & Q-TRACE & QBER/basis & QBER pooled \\
\midrule
Honest (no probe) & 66 & 1.09 & 0.92 & 0.95 & 0.92 \\
\midrule
$\theta=0.15$ & 14 & 1.48 & 1.00 & 1.00 & 1.00 \\
$\theta=0.25$ & 13 & 2.06 & 1.00 & 0.92 & 1.00 \\
$\theta=0.40$ & 14 & 3.10 & 1.00 & 1.00 & 1.00 \\
$\theta=0.60$ & 14 & 5.53 & 1.00 & 1.00 & 1.00 \\
$\theta=0.90$ & 13 & 10.13 & 1.00 & 1.00 & 1.00 \\
$\theta=1.57$ & 13 & 24.17 & 1.00 & 1.00 & 1.00 \\
\midrule
$\gamma=0.03$ & 13 & 4.06 & 1.00 & 1.00 & 1.00 \\
$\gamma=0.06$ & 12 & 4.47 & 1.00 & 1.00 & 1.00 \\
$\gamma=0.12$ & 12 & 7.15 & 1.00 & 1.00 & 1.00 \\
$\gamma=0.25$ & 14 & 11.06 & 1.00 & 1.00 & 1.00 \\
\bottomrule
\end{tabular}

\end{table}

\begin{table}[!htbp]
\centering
\caption{Condition-matched reference: shots per circuit at which half of the attacked link-epochs are flagged at $\alpha=10^{-3}$ (left; 50 is the smallest value tested), and information gathered by Eve before that point, in bits of sifted key (right).}
\label{tab:hwleak}
\small
\begin{tabular}{lcccccc}
\toprule
 & \multicolumn{3}{c}{Shots per circuit to detection} & \multicolumn{3}{c}{Bits leaked before alarm} \\
\cmidrule(lr){2-4}\cmidrule(lr){5-7}
Probe & Q-TRACE & QBER/basis & QBER pooled & Q-TRACE & QBER/basis & QBER pooled \\
\midrule
$\theta=0.15$ & 511 & 432 & 2851 & 14.9 & 12.6 & 83.0 \\
$\theta=0.25$ & 340 & 369 & 694 & 19.7 & 21.3 & 40.1 \\
$\theta=0.40$ & 73 & 69 & 88 & 13.2 & 12.5 & 15.8 \\
$\theta=0.60$ & 50 & 50 & 50 & 18.2 & 18.2 & 18.2 \\
$\theta=0.90$ & 50 & 50 & 50 & 39.0 & 39.0 & 39.0 \\
$\theta=1.57$ & 50 & 50 & 50 & 69.9 & 69.9 & 69.9 \\
\midrule
$\gamma=0.03$ & 60 & 50 & 76 & 0.4 & 0.4 & 0.5 \\
$\gamma=0.06$ & 50 & 50 & 50 & 1.1 & 1.1 & 1.1 \\
$\gamma=0.12$ & 50 & 50 & 50 & 1.9 & 1.9 & 1.9 \\
$\gamma=0.25$ & 50 & 50 & 50 & 8.3 & 8.3 & 8.3 \\
\bottomrule
\end{tabular}

\end{table}

\begin{table}[!htbp]
\centering
\caption{Eve's information measured on her ancilla. $I_{\rm match}$: mutual information per sifted round in which her probe basis matched Alice's basis (Miller--Madow corrected \cite{Miller1955}); $P_{\rm guess\,err}$: her guessing error on those rounds; $I_{\rm sift}$: information per sifted round averaged over all sifted rounds.}
\label{tab:hwinfo}
\small
\begin{tabular}{lcccc}
\toprule
Probe & $I_{\rm match}$ (meas.) & $I_{\rm match}$ (ideal) & $P_{\rm guess\,err}$ & $I_{\rm sift}$ (meas.) \\
\midrule
$\theta=0.15$ & 0.015 & 0.016 & 0.424 & 0.007 \\
$\theta=0.25$ & 0.029 & 0.045 & 0.399 & 0.014 \\
$\theta=0.40$ & 0.090 & 0.112 & 0.324 & 0.045 \\
$\theta=0.60$ & 0.182 & 0.244 & 0.255 & 0.091 \\
$\theta=0.90$ & 0.390 & 0.505 & 0.148 & 0.195 \\
$\theta=1.57$ & 0.699 & 1.000 & 0.053 & 0.349 \\
\midrule
$\gamma=0.03$ & 0.004 & 0.015 & 0.484 & 0.002 \\
$\gamma=0.06$ & 0.011 & 0.031 & 0.471 & 0.006 \\
$\gamma=0.12$ & 0.019 & 0.063 & 0.450 & 0.010 \\
$\gamma=0.25$ & 0.083 & 0.138 & 0.386 & 0.041 \\
\bottomrule
\end{tabular}

\end{table}

\FloatBarrier
\section{Reproducibility}\label{app:repro}

All results were produced with Python 3.12, NumPy 2.5, SciPy 1.18, pandas 3.0, scikit-learn 1.9, XGBoost 3.4, Qiskit 2.5.2, Qiskit Aer 0.17.2 and Qiskit IBM Runtime 0.49.0. The repository provides \texttt{make benchmark} (all held-out seeds), \texttt{make rehearsal} (noisy simulation of the hardware experiment), \texttt{make figures} (all tables and figures) and the notebook \texttt{notebooks/01\_qpu\_hardware\_run.ipynb} used for the hardware run, including its budget control: a pilot job measures the cost per shot and the main job is sized to a user-set limit. Every random draw in the benchmark derives from the seed, so a seed reproduces its result files exactly with the pinned package versions. The numbers in this paper are generated from the result files by the script \texttt{paper/build\_numbers.py}.

\end{document}